\documentclass[twocolumn]{aastex6}

\shorttitle{Sigma Orionis}
\shortauthors{Schaefer et al.}

\begin{document}

\title{Orbits, Distance, and Stellar Masses of the Massive Triple Star $\sigma$ Orionis}

\author{G. H. Schaefer\altaffilmark{1}, C. A. Hummel\altaffilmark{2}, D. R. Gies\altaffilmark{3}, R. T. Zavala\altaffilmark{4}, J. D. Monnier\altaffilmark{5}, F. M. Walter\altaffilmark{6}, N. H. Turner\altaffilmark{1}, \\ F. Baron\altaffilmark{3}, T. ten Brummelaar\altaffilmark{1}, X. Che\altaffilmark{5}, C. D. Farrington\altaffilmark{1},  S. Kraus\altaffilmark{7}, J. Sturmann\altaffilmark{1}, and L. Sturmann\altaffilmark{1}}

\altaffiltext{1}{The CHARA Array of Georgia State University, Mount Wilson Observatory, Mount Wilson, CA 91023, U.S.A.; schaefer@chara-array.org}
\altaffiltext{2}{European Southern Observatory, Karl-Schwarzschild-Str.~2, 85748 Garching, Germany; chummel@eso.org}
\altaffiltext{3}{Center for High Angular Resolution Astronomy and Department of Physics and Astronomy, Georgia State University, P.O. Box 5060, Atlanta, GA 30302-5060, USA; gies@chara.gsu.edu}
\altaffiltext{4}{U.S. Naval Observatory, Flagstaff Station, 10391 W. Naval Obs. Rd., Flagstaff, AZ 86001, USA; bzavala@nofs.navy.mil}
\altaffiltext{5}{Department of Astronomy, University of Michigan, Ann Arbor, MI 48109, USA}
\altaffiltext{6}{Department of Physics and Astronomy, Stony Brook University, Stony Brook, NY 11794-3800, USA}
\altaffiltext{7}{Astrophysics Group, School of Physics, University of Exeter, Stocker Road, Exeter, EX4 4QL, UK}

\begin{abstract}
We present interferometric observations of the $\sigma$ Orionis triple system using the CHARA Array, NPOI, and VLTI.  Using these measurements, we spatially resolve the orbit of the close spectroscopic binary (Aa,Ab) for the first time and present a revised orbit for the wide pair (A,B).  Combining the visual orbits with previously published radial velocity measurements and new radial velocities measured at CTIO, we derive dynamical masses for the three massive stars in the system of $M_{\rm Aa}$ = 16.99  $\pm$ 0.20 $M_\odot$, $M_{\rm Ab}$ = 12.81  $\pm$ 0.18 $M_\odot$, and $M_{\rm B}$ = 11.5 $\pm$ 1.2 $M_\odot$.  The inner and outer orbits in the triple are not coplanar, with a relative inclination of 120$^\circ$--127$^\circ$.  The orbital parallax provides a precise distance of 387.5 $\pm$ 1.3 pc to the system.  This is a significant improvement over previous estimates of the distance to the young $\sigma$ Orionis cluster.
\end{abstract}

\keywords{binaries: spectroscopic --- binaries: visual ---
          stars: fundamental parameters --- stars: individual ($\sigma$ Orionis) ---
          techniques: interferometric
          } 

\section{Introduction}

The $\sigma$ Orionis cluster contains several hundred young stars surrounding the multiple star system $\sigma$ Orionis \citep[see review by][]{walter08}.  The clustering of 15 B-type stars in the region was first noted by \citet{garrison67}, and it was included in the catalog of open clusters by \citet{lynga81}.  The discovery of a large population of low-mass pre-main sequence stars in the area around $\sigma$ Orionis was reported by \citet{walter97,walter98}.  Subsequent photometric and spectroscopic searches have identified additional low-mass and substellar candidate members \citep[e.g.,][]{bejar99, bejar11, zapatero00, sherry04, caballero08, lodieu09, hernandez14, koenig15}.  With an age of about 2$-$3 Myr \citep{sherry08}, about 30\% to 50\% of low-mass stars ($M < 1$ $M_\odot$) in the cluster retain their accretion disks \citep[e.g.,][]{oliveira06,hernandez07,luhman08,sacco08,pena_ramirez12}.  Distance estimates to the cluster range from 330 to 450 pc \citep[e.g.,][]{walter08}.

The multiple star system $\sigma$ Orionis (HD 37468, WDS J05387-0236) lies at the center of the cluster.  The five main components include the O9~V star $\sigma$ Ori A, the B0.5~V star $\sigma$ Ori B at a separation of 0\farcs25 \citep{burnham1894,edwards76}, the A2~V star $\sigma$ Ori C at 11$''$, the B2~V star $\sigma$ Ori D at 13$''$, and the helium-rich, magnetic B2~Vpe star $\sigma$ Ori E at 42$''$ \citep{struve1837,greenstein58,landstreet78}.  A more extensive description of the multiplicity of wider or fainter components in the system is described by \citet{caballero14}.  The pair $\sigma$ Ori A,B has an orbital period of about 157 yr \citep{heintz74,heintz97,hartkopf96,turner08}.  The A component was suspected to be a spectroscopic binary based on the appearance of double lines in the spectrum \citep{frost1904,miczaika50,bolton74}, but it was not until recently that a double-lined spectroscopic binary orbit was measured; the spectroscopic pair $\sigma$ Ori Aa,Ab has a period of 143 days \citep{simondiaz11,simondiaz15}.  

In this paper we report spatially resolved measurements of the close triple system ($\sigma$ Ori Aa,Ab,B) using long baseline optical/infrared interferometry and also present new spectroscopic radial velocity measurements.  Combining the visual orbits with the new and previously published radial velocities yields the dynamical masses of the three components and the distance to the system.  Precise dynamical masses of O-stars are needed for testing the predictions from different sets of evolutionary models for massive stars \citep{maeder95,gies03,weidner10,massey12,morrell14}.  Additionally, a precise orbital parallax to the $\sigma$ Orionis cluster provides an accurate distance for determining the age of the cluster and characterizing the physical properties and disk life-times for the stars, brown dwarfs, and planetary-mass members in the region.

\section{Interferometric Observations of the $\sigma$ Orionis Triple}

A general overview of optical interferometry and measures of the interference fringes (visibility amplitude and closure phase) can be found in reviews on the subject \citep{lawson00,monnier03,haniff07}.  The visibility amplitudes provide information on the size, shape, and structure of the source.  The closure phases are particularly sensitive to asymmetries in the light distribution.

\subsection{CHARA Observations and Data Reduction}

Interferometric data on the $\sigma$ Orionis triple system were collected between 2010 and 2013 at the CHARA Array located on Mount Wilson, California.  The array has six 1\,m telescopes arranged in a $Y$ configuration with baselines ranging from 34 to 331 m \citep{tenbrummelaar05}.  There are two telescopes in each arm, labeled as E (East), W (West), and S (South).  We used the Michigan Infrared Combiner \citep[MIRC;][]{monnier04,monnier06} to combine the light from three to six telescopes simultaneously.  All data were collected after the photometric channels were installed in MIRC; the photometric channels measure the amount of light received from each telescope during the observations to improve the calibration \citep{che10}.  We used the low spectral resolution prism ($R \sim 42$) to disperse the fringes across eight spectral channels in the $H$-band ($\lambda = 1.5-1.8 \mu$m).  Table~\ref{tab.log} provides an observing log that lists the UT date, HJD, telescope configuration, interferometric calibrator stars used during the observations, the number of visibility and closure phase measurements recorded on each night, and the median seeing corrected to zenith in the $V$-band reported by the tip-tilt sytem during the $\sigma$ Orionis observations.

\begin{figure*}
  \plotone{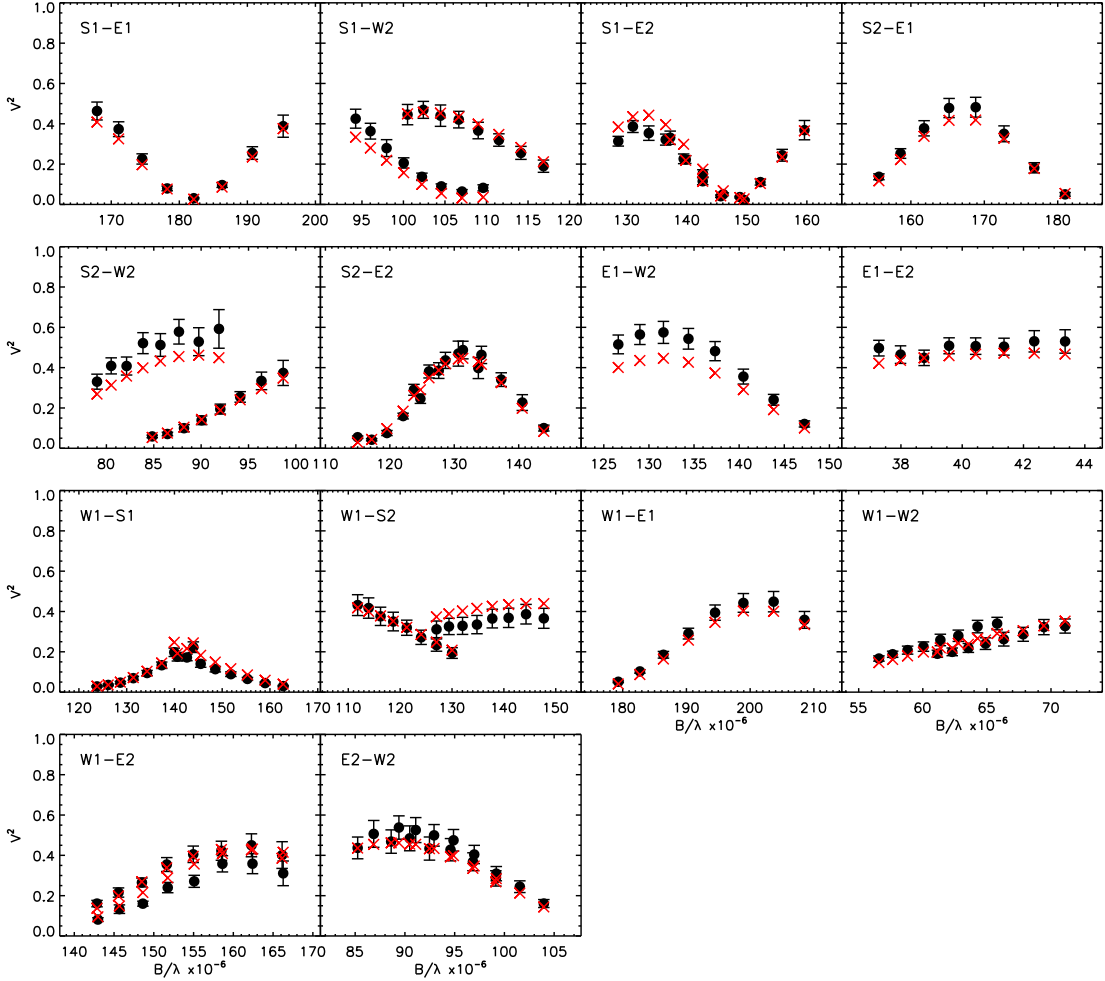}
  \caption{Squared visibilities of $\sigma$ Orionis measured with MIRC at the CHARA Array on UT 2011 September 29 (filled black circles).  The red crosses indicate the visibilities derived from the best-fit scaled binary model.  The observations have been averaged over 5 min observing blocks.  The S1-S2 baseline has been excluded from the fit (see text). }
\label{fig.vis2_bin}
\end{figure*}

\begin{figure*}
  \plotone{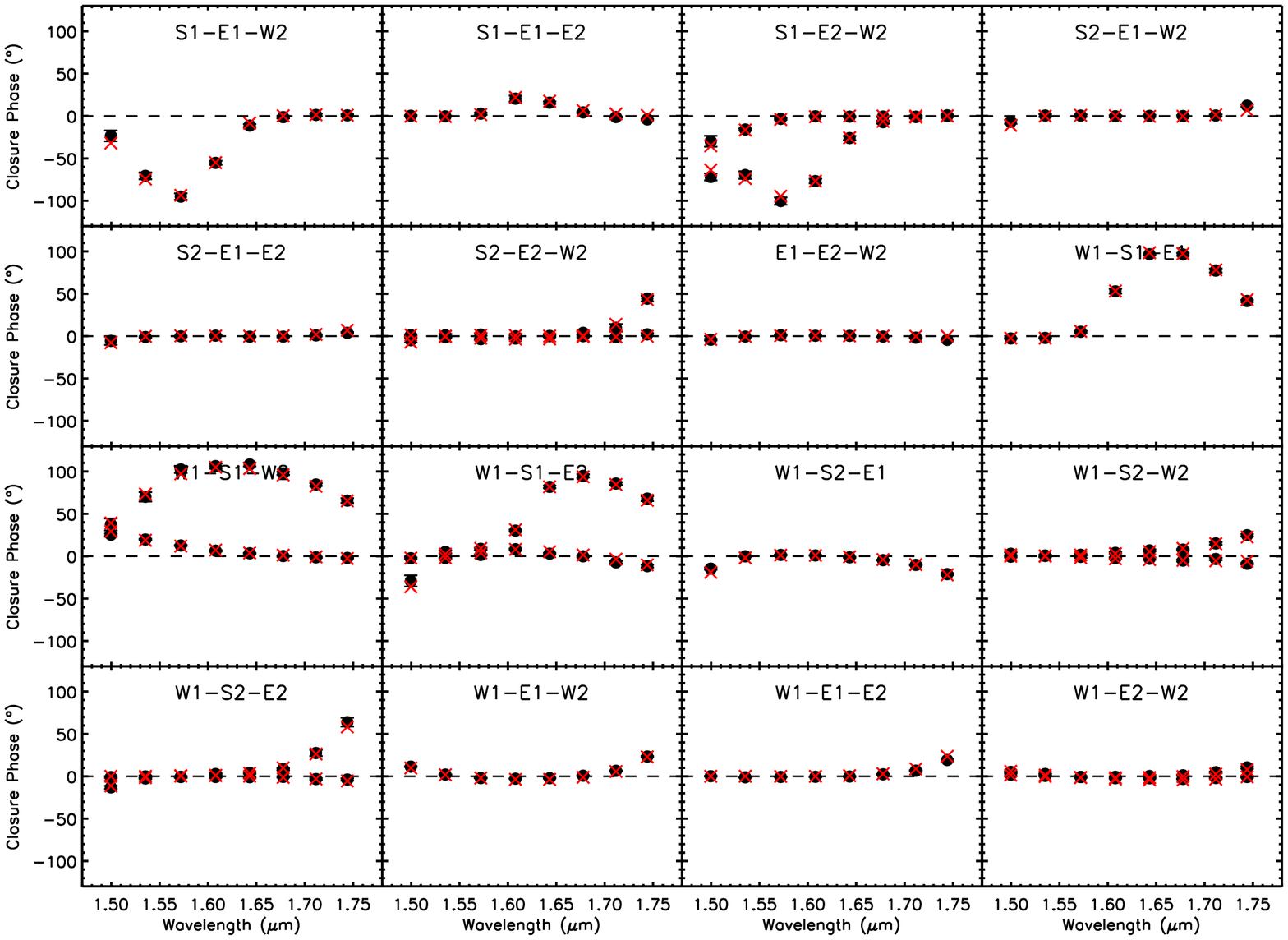}
  \caption{Closure phases of $\sigma$ Orionis  measured with MIRC at the CHARA Array on UT 2011 September 29 (filled black circles).  The red crosses indicate the closure phases derived from the best-fit scaled binary model.  The observations have been averaged over 5 min observing blocks.  Closure triangles that include the S1-S2 baseline have been excluded from the fit (see text). }
\label{fig.t3_bin}
\end{figure*}

The CHARA data were reduced using the standard MIRC reduction pipeline \citep[e.g.,][]{monnier07}.  For nearly all nights, we used a coherent integration time of 75 ms to improve the signal to noise.  On UT 2010 November 5, we found differences in the visibility calibration using the 75 ms coherent integration time compared with the default value of 17 ms; this was probably because of rapid time variability in the seeing.  For that night, we used the squared visibilities from the 17 ms integration times and the closure phases from the 75 ms integration times.  The data were calibrated using observations of single stars of known angular sizes observed before and/or after the target.   The adopted angular diameters for the calibrator stars are listed in Table~\ref{tab.cal}.  For HD 25490 and HD 33256 we computed the angular diameters by modeling their spectral energy distributions using the method described in \citet{schaefer10}.  The calibrated data were averaged over 5$-$30 minute observing blocks.  Based on calibrator studies, we applied minimum uncertainties of 5\% on the squared visibilities and 0\fdg3 on the closure phases.  We corrected the wavelength scale according to the wavelength calibration computed by \citet{monnier12}.  The precision in the absolute wavelength calibration is good to $\pm$0.25\%.  Examples of the calibrated squared visibility amplitudes and closure phases of $\sigma$ Orionis are shown in Figures~\ref{fig.vis2_bin} and~\ref{fig.t3_bin}.  The calibrated data files with the systematic uncertainties and wavelength correction applied will be available through the Optical Interferometry Database developed by the Jean-Marie Mariotti Center\footnote{http://www.jmmc.fr/oidb.htm}.

\subsection{CHARA Astrometric Results}

The diffraction limit of a single 1\,m CHARA telescope in the $H$-band corresponds to $\sim$ 0\farcs4 on the sky.  Therefore, light from all three components in the $\sigma$ Orionis triple (Aa, Ab, B) is recorded in the field of view of the detector (set by the injection of light into the optical fibers of MIRC).  However, given the width of the MIRC spectral channels ($\Delta \lambda \sim 0.035$~$\mu$m) and the corresponding coherence length ($\lambda^2/\Delta\lambda \sim 75$~$\mu$m), the wide 0\farcs25 component $\sigma$ Ori B contributes only incoherent light on all but the shortest baselines, degrading the fringe amplitude by a constant amount set by the percentage of light coming from the wide component.  For the shortest baselines (e.g., S1-S2 with a baseline length of 34~m), light from the wide component adds coherently to produce additional periodic variations in the visibilities and closure phases.  To simplify the model fitting, we excluded from the fit the S1-S2 baseline and all closure triangles that included both the S1 and S2 telescopes.

A binary star produces a periodic signal in the complex fringe visibilities \citep{boden00}.  The presence of the wide third component adds incoherent flux that can be accounted for by scaling the complex visibilities,
\begin{equation}
V = \frac{f_1V_1 + f_2V_2 \exp{[-2\pi i (u\Delta \alpha+ v\Delta \delta)]}}{(f_1 + f_2 + f_3)}
\end{equation}
where ($\Delta \alpha, \Delta \delta$) are the close pair binary separation in R.A. and Decl., ($u,v$) are the baseline components projected on the sky, $V_1$ and $V_2$ are the uniform disk visibilities of the primary and secondary components with angular diameters $\theta_1$ and $\theta_2$, and $f_1$, $f_2$, and $f_3$ are the flux fractions from each of the three components ($f_1 + f_2 + f_3 = 1$).   When $f_3$ is non-zero, the peaks in the periodic visibility curves no longer rise to 1.  The real and imaginary parts of the complex visibility are combined to form the squared visibility amplitude between each pair of telescopes and the closure phase for each set of three telescopes.  We fit the squared visibility amplitudes and closure phases measured with MIRC using this scaled binary model, assuming angular diameters for the component stars of $\theta_{\rm Aa} = 0.27$ mas and $\theta_{\rm Ab}$ = 0.21 mas (see Section~\ref{sect.npoi_orb}).  The adopted values are larger than the angular diameters predicted by \citet[][0.14 mas and 0.12 mas]{simondiaz15}.  However, because the stellar diameters are unresolved by the interferometer, the effect on the model fitting is small.  The flux contributions change by about 1$-$2\%, while the binary positions remain consistent within the 1\,$\sigma$ uncertainties.

We followed an adaptive grid search procedure \citep[similar to the method described in][]{gallenne15} where we searched through a grid of separations in R.A. and Decl. and performed a Levenberg-Marquardt least-squares minimization using the IDL mpfit\footnote{http://cow.physics.wisc.edu/\raisebox{0.1em}{\tiny$\sim$\,}craigm/idl/idl.html} routine \citep{markwardt09} to determine the best fit binary solution for each step in the grid.  We retained the solution with the lowest $\chi^2$ and examined the $\chi^2$ space to check for possible alternative solutions.  For most epochs we found a unique solution with a second minimum reflected through the origin but with the fluxes of the components in the close pair flipped (no other solutions were typically found within $\Delta\chi^2 > 100-10,000$ from the best fit).  For the data taken on UT 2013 November 3, we found an alternative solution with $\Delta\chi^2$ = 12 from the best fit solution; in addition to the higher $\chi^2$, the alternative position is not consistent with the orbital motion mapped in Section~\ref{sect.orbAaAb}.  On UT 2010 November 4, we found multiple solutions in the $\chi^2$ maps with $\chi^2 < 25$.  This was likely caused by a combination of the limited $(u,v)$ coverage during the observation and poor data calibration because of possible alignment drifts during the long time interval to find fringes combined with poor seeing conditions as the target was setting (altitude $\sim$ 36$^\circ$).  Because of the ambiguities in the solutions, we do not report a position for this night.

Table~\ref{tab.sepPA_mirc} lists the separation $\rho$, position angle $\theta$ (measured east of north), and component flux contributions during each of the MIRC observations obtained at the CHARA Array.  Uncertainties in the binary positions were computed from the covariance matrix and include correlations between the binary separation in R.A. and Decl.  In Table~\ref{tab.sepPA_mirc}, we report the semi-major axis, semi-minor axis, and position angle of the major axis of the error ellipse ($\sigma_{\rm maj}, \sigma_{\rm min}, \phi$, respectively).  We compared these uncertainties against $\chi^2$ maps generated from a two-dimensional grid search using fixed steps in separation; the error ellipses are in agreement with the size and orientation of the 1\,$\sigma$ ($\Delta \chi^2 = 1$) confidence intervals from the $\chi^2$ maps.  On average, the components contribute a mean of 47.7\% $\pm$ 5.9\% (Aa), 27.4\% $\pm$ 5.2\% (Ab), and 24.9\% $\pm$ 8.0\% (B) of the total light recorded on the detector in the $H$-band.  These fractional flux contributions are very similar to those in the $V$-band as estimated by \citet{simondiaz15}, $48\% $, $28\% $ and $24\% $, respectively.

The larger uncertainties derived for the binary positions on 2013 November 3 and 11 are likely caused by a combination of poor seeing conditions that made finding and tracking the fringes difficult, and the limited $(u,v)$ coverage obtained from the smaller number of telescopes on which fringes could be found.  The binary position is expected to change more rapidly on these nights since the companion is near periastron, however, the expected motion on the sky during the time-frame of the observations is smaller than the measurement uncertainties.  Breaking the data into smaller time blocks that were fit independently resulted in positions that varied randomly with even bigger error ellipses.  Therefore we report the average positions based on the fit to all measurements on each night.

As a check on our results, we also fit the MIRC data using a triple model that includes the relative separation between all three components, $\sigma$ Ori Aa,Ab,B.  To minimize the effects of time smearing, we used calibrated data files that were averaged over shorter 2.5 min observing blocks.  To account for time smearing across the observing blocks, we computed the triple model at 10 second intervals and averaged over the complex visibilities.  We also accounted for bandwidth smearing which reduces the fringe coherence at separations comparable to the width of the fringe packet following the formalism in \citet{kraus05}.  Summing the visibilities at the location of each component, the complex visibility of a triple system is given by,
\begin{eqnarray}
V &=& \big[ f_1 c_1(\tau) V_1 e^{-2 \pi i (u \Delta \alpha_{1} + v \Delta \delta_{1})}  \nonumber \\ 
  & & + f_2 c_2(\tau) V_2 e^{-2 \pi i (u \Delta \alpha_{2} + v \Delta \delta_{2})} \nonumber \\ 
  & & + f_3 c_3(\tau) V_3 e^{-2 \pi i (u \Delta \alpha_{3} + v \Delta \delta_{3})} \big] \times \frac{1}{f_1 + f_2 + f_3} ~~~~~~
\end{eqnarray}
where $(\Delta \alpha_{n}, \Delta \delta_{n})$ are the separations in R.A. and Decl.\ between the primary, secondary, and tertiary components ($n$ = 1, 2, 3) and the phase center.  In the analysis of the MIRC data, we assumed the phase center to be the photocenter of $\sigma$ Ori Aa,Ab.  The coherence for a rectangular bandpass profile is given by
\begin{equation}
c_n(\tau) = \frac{\sin{(\pi \tau_{n} \Delta\lambda/\lambda^2)}}{\pi \tau_{n} \Delta\lambda/\lambda^2} 
\end{equation}
where the optical path length delays are given by
\begin{equation}
\tau_{n} = \lambda (u \Delta \alpha_{n} + v \Delta \delta_{n})
\end{equation}
and $\Delta \lambda$ is the width of the wavelength channel and $\lambda$ is the central wavelength.

\begin{figure*}
  \scalebox{0.55}{\includegraphics{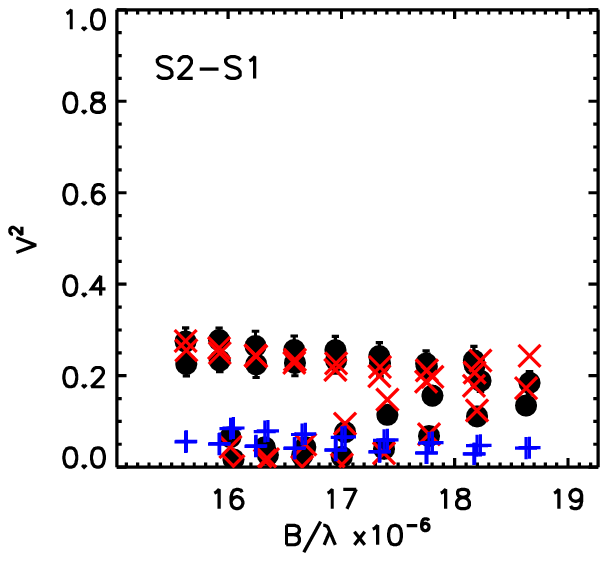}}
  \scalebox{0.55}{\includegraphics{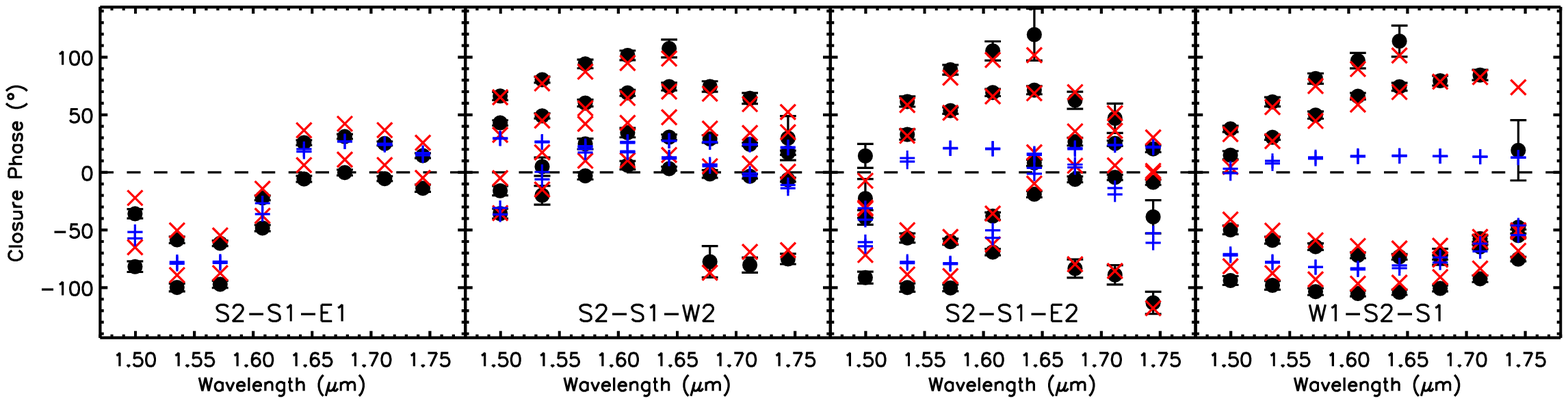}}
  \caption{Squared visibilities (left) and closure phases (right) of $\sigma$ Ori  measured with MIRC at the CHARA Array on UT 2011 September 29 (black circles).  We show only the shortest baseline and closure triangles that include the S1-S2 telescopes.  The observations have been averaged over 2.5 min observing blocks.  The blue plus signs show the best-fit scaled binary model fit to all baselines and triangles.  The observations obtained with the shortest baseline are fit much better by a triple model (red crosses) that directly includes the position of the wide companion $\sigma$ Ori B.}
\label{fig.triple}
\end{figure*}

The triple model reproduces the variation in the visibilities and closure phases on the baselines and triangles that include the S1 and S2 telescopes as shown in Figure~\ref{fig.triple}.  However, the triple fit is further complicated by changes in seeing and telescope-dependent tip-tilt corrections that influence the measured photocenter of the system and the corresponding phase shift of the fringes.  The wide component is over-resolved on the longer baselines, so it is primarily the short S1-S2 baseline that samples the wide pair separation.  Because of this limited baseline coverage on the sky, the $\chi^2$ maps for the wide component separation sometimes have multiple peaks that are consistent with the data.  On the other hand, the close pair separations derived from the triple model are stable and within the uncertainties of those from the scaled binary fit.  We opted to report the simpler scaled binary solution as our final results.  

\subsection{VLTI Observations and Data Reduction}

$\sigma$ Orionis was observed with the AMBER \citep{petrov07}
beam combiner at the VLTI \citep{scholler07} using the Antu (UT1), 
Kueyen (UT2) and Yepun (UT4) 8.2\,m telescopes on UT 2008 October 14
(HJD 2454753.7).  The data were recorded with the low-resolution mode ($R=35$)
in the $H$ and $K$ bands. The longest baseline between UT1 and UT4 is
nominally 130 m in length. A single observation of the science target
was sandwiched between two calibrator observations, one of HD 34137
and the other of HD 36059, with diameters of $0.73 \pm 0.02$ mas and 
$0.51 \pm 0.01$ mas, respectively \citep{bonneau06,bonneau11}.  The data
were reduced using the amdlib pipeline \citep{tatulli07,chelli09}
but only the top 30\% visibility data in terms of signal-to-noise ratio
were used to reduce the influence of periods of poor group-delay fringe
tracking.  Seeing was 0\farcs8 on average, but vibrations present in the UT
infrastructure limited the fringe contrast.  The transfer function was
linearly interpolated between the two calibrator measurements to the
epoch of the science observation.

\begin{figure}
\plottwo{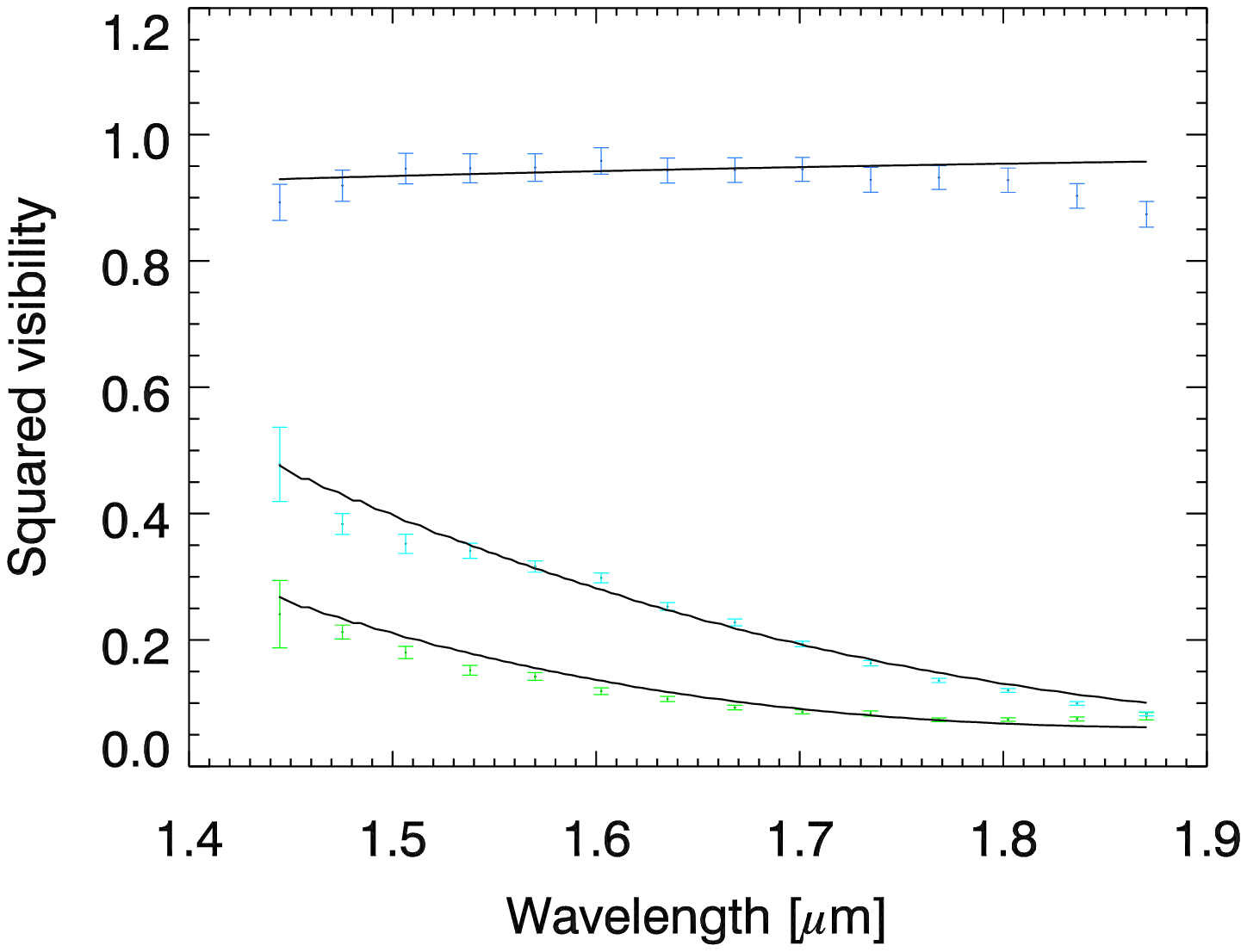}{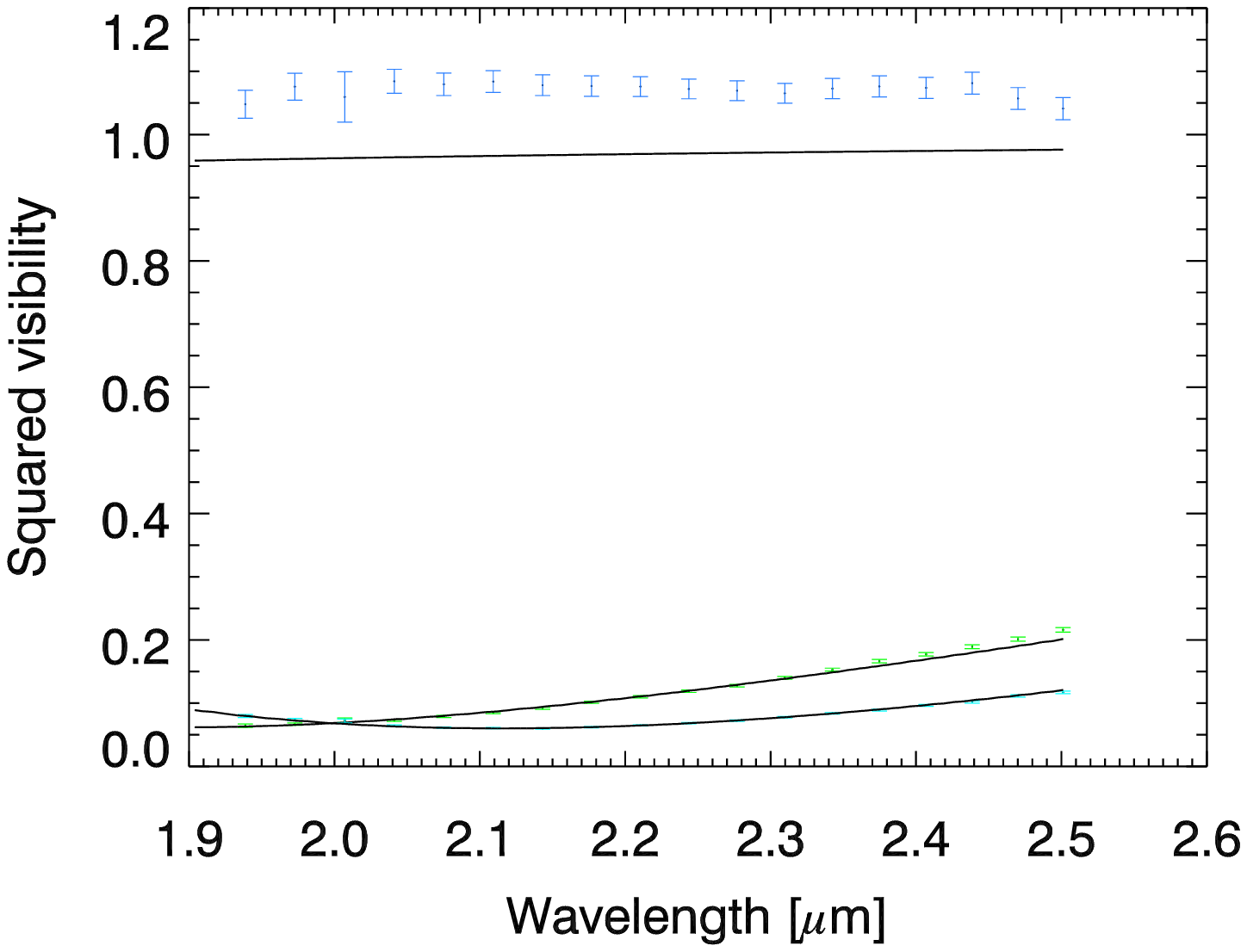}\\
\plottwo{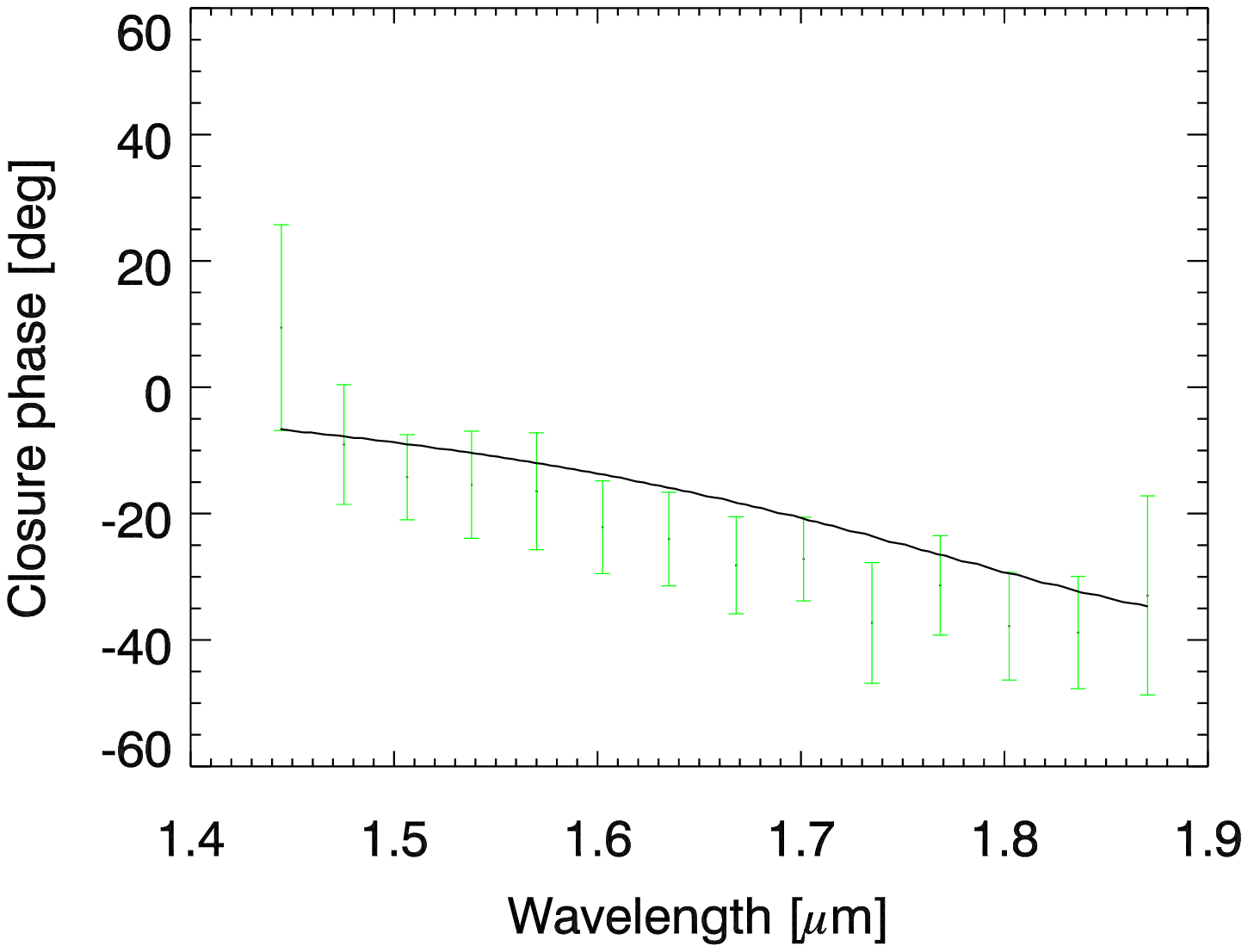}{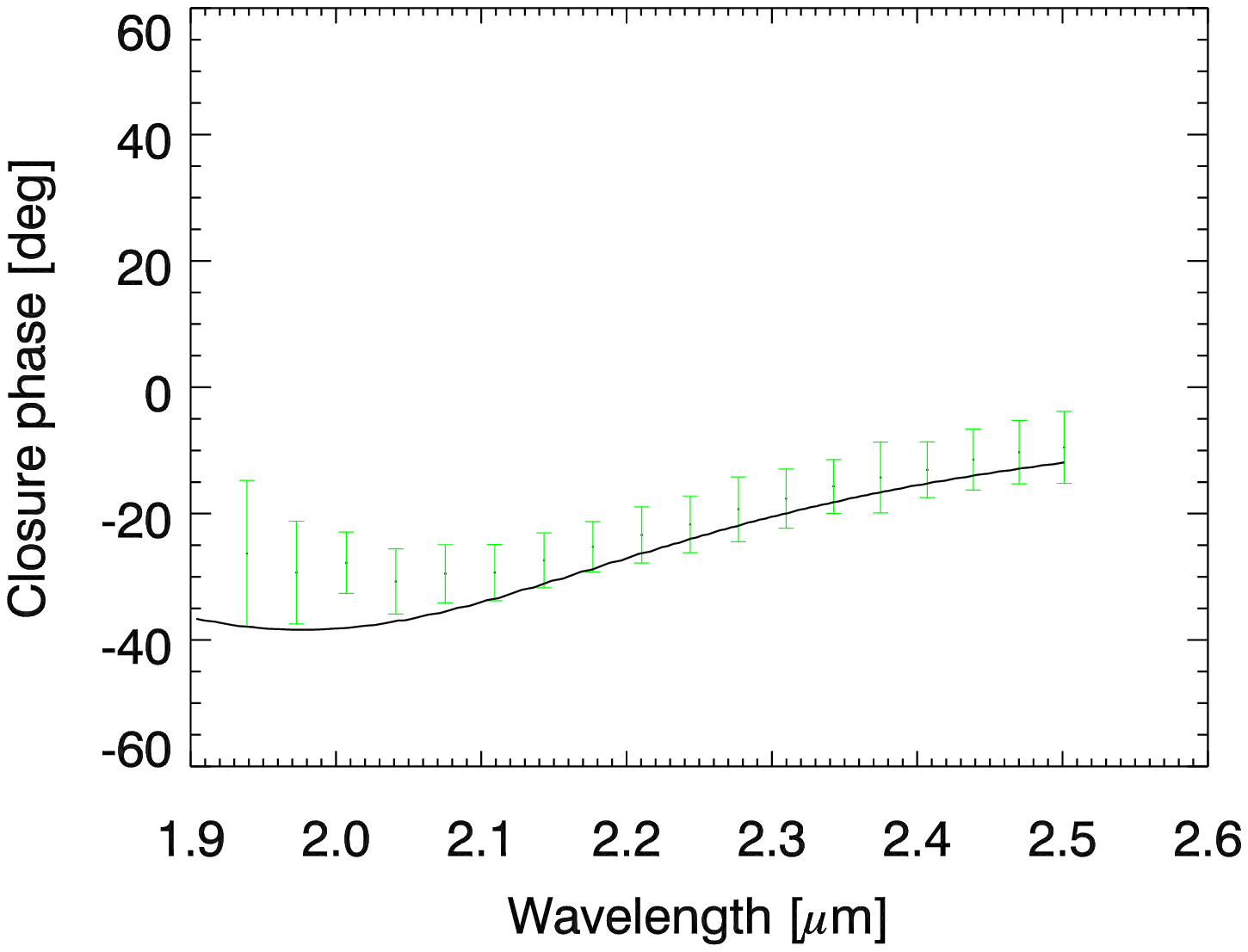}\\
\caption{AMBER $H$ and $K$ (squared) visibilities and closure phases.  Each color in the visibility plots represents a different baseline pair.  The best-fit binary model is overplotted as the solid black line.}
\label{AMBER_fig_orbit}
\end{figure}

With a field of view of about 60 mas with the UTs, AMBER only sees the close 
pair.  The measured separation and position angle are $\rho=4.30\pm0.52$ mas
and $\theta = 174\fdg70 \pm 4\fdg7$. The fit to the data is shown in
Fig.~\ref{AMBER_fig_orbit}. The fitted magnitude difference between Ab and
Aa is $0.57\pm0.03$ in the $H$ band and $0.55\pm0.02$ in the $K$ band. The
$H$ band value is consistent with the CHARA value of $\Delta H=0.60$ mag.

\subsection{NPOI Observations and Data Reduction}

NPOI observations \citep{armstrong98} of $\sigma$ Orionis were collected 
over a period from 2000 to 2013.  Initially, the observations were obtained 
with the 3-beam combiner, and 
then, starting in 2002, with the 6-beam hybrid combiner \citep{benson03}.
The NPOI beam combiners disperse the light
and record the visibility spectra from 550 nm to 850 nm in 16 
spectral channels. In total, some 59 nights of
observations were executed, of which 26 nights were of good quality.
Observations of calibrator stars were interleaved with the science
target. Table~\ref{NPOI_table_obs} gives information on dates,
configurations, and calibrator stars observed for each night.
A configuration is given as a triple of stations (e.g. ``AC-AE-W7'',
using astrometric stations Center and East, as well as imaging station
W7) if data from all three baselines were used, including the
corresponding closure phase. If a single baseline is listed, squared
visibility data from that baseline were used but no closure phase data
were available involving this baseline.

The calibrators were selected from a list of single stars maintained at
NPOI with diameters estimated from $V$ and $(V-K)$ using the surface
brightness relation published by \citet{mozurkewich03} and
\citet{vanbelle09}.  Estimates for $E(B-V)$ derived by
comparison of the observed colors to theoretical colors as a function of
spectral type as given by Schmidt-Kaler in \citet{aller82} were used to 
derive extinction estimates $A_V$. These were compared to measurements
based on maps by \citet{drimmel03} and used to correct $V$ if the two 
estimates agreed within 0.5 magnitudes.  Even though the surface brightness
relationship based on $(V-K)$ colors is to first order independent
of reddening, we included this small correction because our principal
calibrator, $\epsilon$ Orionis (HD 37128), is a B-supergiant at more than 
400 pc distance and has a predicted apparent diameter of 1.01 mas.  
Based on an analysis of calibrator stars observed using the Mark III 
interferometer, \citet{mozurkewich91} measured uniform disk diameters of 
0.86 mas $\pm$ 0.16 mas (at 800 nm) and 1.02 $\pm$ 0.12 mas (at 450 nm) for 
$\epsilon$ Orionis.  However, because the star was barely resolved on the 
Mark III baselines (up to 38 m in length), we decided to use our estimate as 
the more precise value.  On the longest NPOI baseline that we used 
(E6-W7, 79 m), and in the middle of the bandpass (700 nm), the expected 
squared visibility of $\epsilon$ Orionis is 0.45.  The information for all 
of the calibrators is given in Table~\ref{NPOI_table_cal}.

The NPOI data and their reduction were described by \citet{hummel98,hummel03}. 
We used an new version of the OYSTER\footnote{http://www.eso.org/$\sim$chummel/oyster} 
NPOI data reduction package written in GDL\footnote{http://gnudatalanguage.sourceforge.net}.  
The pipeline automatically edits the 1-second averages
produced by another pipeline directly from the raw frames, based on
expected performance such as the variance of fringe tracker delay, photon
count rates, and narrow angle tracker offsets. Visibility bias corrections
are derived as usual from data recorded away from the stellar fringe
packet. After averaging the data over the full length of an observation,
the closure phases and the transfer function of the calibrators
were interpolated to the observation epochs of $\sigma$ Orionis.  For the
calibration of the visibilities, the pipeline used all calibrator stars
observed during a night to obtain smooth averages of the amplitude
and phase transfer functions using a Gaussian kernel of 80 minutes
in length.  The residual scatter of the calibrator visibilities and
phases around the average set the level of the calibration uncertainty
and was added in quadrature to the intrinsic data errors. 

Considerable effort was invested in algorithms that automatically edit the visibility data based on the variance of the delay-line positions following the procedures described by \citet[][Section 4.2]{hummel03} and adapted to more complicated source structures where the signal-to-noise ratio is low.
Especially in the case of $\sigma$ Orionis, deep visibility minima exist on the 
baselines typically employed by our observations. A final step was therfore added
to detect problems by comparing the results to the predictions of the
final model derived later from all data sets.  An amplitude
calibration error of typically a few percent in the red channels and up to
15\% in the blue channels was added in quadrature to the intrinsic error
of the visibilities. The phase calibration was good to $\sim 2^\circ$.
Nevertheless, because of small changes in atmospheric conditions
between the observations of the calibrators and the science target we
used additional {\em baseline-based} calibration factors (``floating 
calibration'') to allow minor adjustments of the visibility spectra to obtain 
better fits to the orbital elements (and magnitude differences) of the triple 
system.  Two thirds of the spectra were adjusted by less than 25\%, the remainder
were mostly low SNR spectra. Because the components of $\sigma$ Orionis Aa, Ab, and B
are unresolved (see Section~\ref{sect.npoi_orb}), the maximum visibility amplitude was
fixed to unity.  This procedure will not bias the astrometric results because the
binary separation is constrained mostly by the variation of the visibility data with 
wavelength (Fig.~\ref{NPOI_fig_vissq325}). The magnitude differences between the 
components across the 550$-$850 nm band were determined to be 0.5 $\pm$ 0.1 mag 
for Ab$-$Aa and 1.5 $\pm$ 0.2 mag for B$-$A.  We assumed that the magnitude 
differences between the components are the same across the $V$ and $I$ bands; 
this is expected since both components are hot stars and should have similar 
colors.

\subsection{NPOI Astrometric Results}
\label{sect.npoi_astrom}

Because of the large angular separation of the tertiary component ($\sigma$ Ori B),
rapid variations of the visibility amplitude occur on the shorter
NPOI baselines, while they are completely smeared out on the longer
baselines due to the finite width of the spectral channels.  The number
of fringes in the central envelope of an interferogram is given by 
$N=2\lambda/\Delta\lambda=2R$ where $\Delta\lambda$ is the width of the 
bandpass and $R$ is the equivalent spectral resolving power of the 
spectrometer. The fringe 
amplitude decreases to zero towards the edge of the envelope.  One fringe
spacing corresponds to $\lambda/B$ radians on the sky, where $B$ is the
projected baseline length.  Since the smallest baselines employed for
our observations are about 20 M$\lambda$ long (in the reddest channel at
850 nm), the fringe spacing is about 10 mas, and thus the field of view
is about 300 mas in diameter if we consider a loss in (squared) amplitude
of about 60\% and $R=30$ for the NPOI spectrometers.  Since the NPOI
channel bandpasses are known, complex visibilities predicted by a model
of the triple system are computed on a sufficiently fine wavelength grid,
and then averaged over the bandpasses before converting them to squared
visibilities and closure phase for comparison to the observed quantities.

\begin{figure*}
\plotone{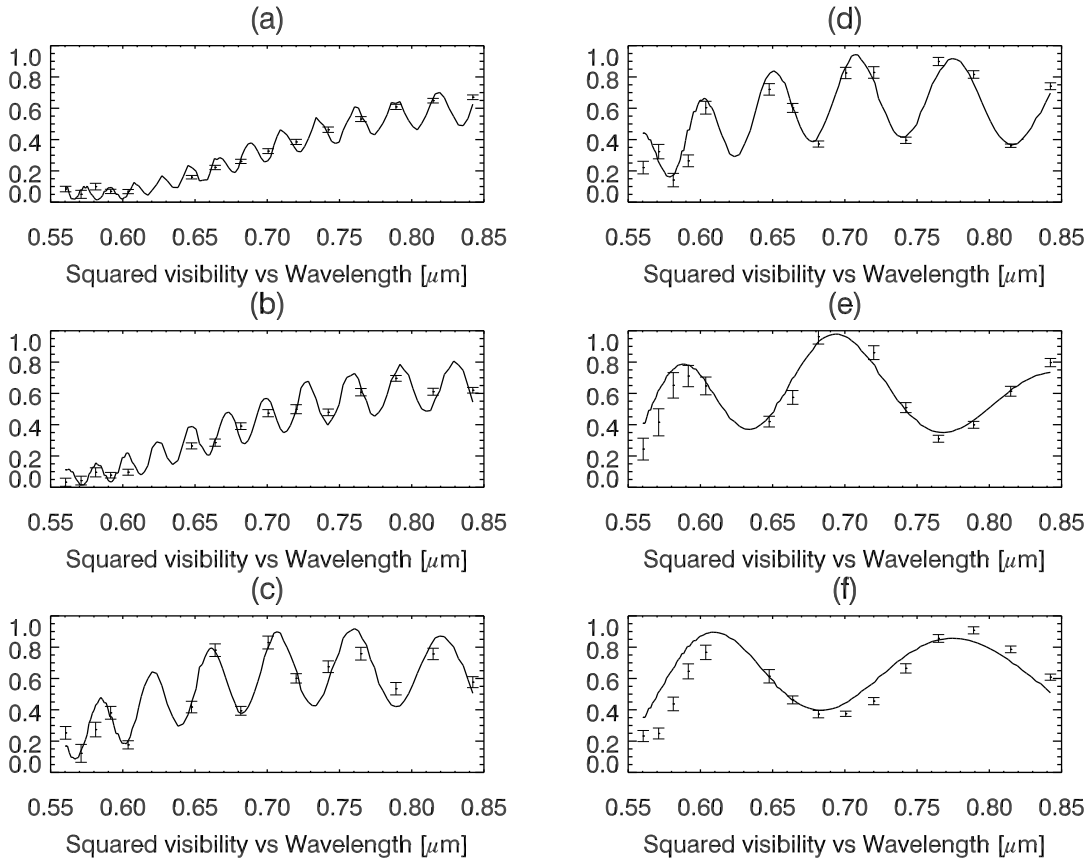}
\caption{NPOI squared visibilities for 2010 March 25. Panels a-f
correspond to observations at 03:38, 03:42, 04:08, 04:17, 04:32, and
04:36 UT.  The data shown are for baseline AN0-W7.
}
\label{NPOI_fig_vissq325}
\end{figure*}

An example of the rapid variations of the (squared) visibility amplitude
on the AN0-W7 baseline is shown in Figure~\ref{NPOI_fig_vissq325},
together with the predicted values from our final model (discussed in
Section~\ref{sect.npoi_orb}). Small errors in the predicted position
of the tertiary relative to the close binary can lead to significant
deviations between the data and the model.  Therefore, we first improved our
knowledge of the tertiary orbit.  The elements published
by \citet{turner08} were based on adaptive optics
and speckle measurements, the last of which dates back to the end of
2001. While our NPOI observations started around the same time, the early
data sets did not allow for the unambiguous identification of the location of
the tertiary (if detected at all) because of the close and regular spacing of
the local minima in the $\chi^2$ surface
which is caused by undersampling of the fast variations of the visibility
amplitude in combination with often parallel orientation of the baselines
relative to the direction of the tertiary. The
first night to provide an unambiguous identification of the position
of the tertiary was on 2010 March 25, as one of the baselines, AN0-W7,
rotated close to an orthogonal orientation to the wide binary orientation,
causing a change in the ``wavelength'' of the visibility oscillation (seen
in Fig.~\ref{NPOI_fig_vissq325}). 
We then added this epoch to the measurements
of $\sigma$ Ori A,B available from the Washington Double Star Catalog
and refit the orbital elements. Subsequently,
five more nights were identified with similar quality, and were used to
refine the orbital elements.  Finally, all nights with a pronounced
minimum of $\chi^2$ at the predicted position of the tertiary were
included in the fit. The results are given in Table~\ref{NPOI_table_abc} which
gives the date, Julian year of the observation (at 7 UT), the number of
measured visibilities, the derived separation (relative to the center of mass
of the close pair), position angle, and the semi-axes and position angle of
the uncertainty ellipses.  The last two columns give the deviation of the 
fitted relative binary position $(\rho,\theta)$ from the model values.  
The uncertainty ellipses were computed from fits to contours 
of the $\chi^2$ surfaces near the minima rather than deriving them from the
interferometric PSF.  This accounts for the limitations of fitting a
component position very far from the phase center.  We scaled the contours
to result in a reduced $\chi^2$ of unity at the minimum.  The positions of 
$\sigma$ Ori A,B are in good agreement with measurements made 
at similar times by \citet{simondiaz15} and \citet{aldoretta15}.

After the orbit of the tertiary was revised,
astrometric positions of the secondary were fit to the visibility data for
each night separately (with fixed tertiary positions derived from the
tertiary orbit). Error ellipses were estimated using the $\chi^2$ 
surface maps centered on the position of the secondary.   The $\chi^2$ 
contour interval was selected to give a reduced $\chi^2$ close to unity 
when fitting the astrometric positions with an orbit for the close pair.
This resulted in using the $\Delta \chi^2 = 40$ confidence interval.
Correlations in the visibility amplitudes between the 16 channels,
related to atmospheric seeing variations, reduces the number of independent 
data points and explains partly the size of this interval.  
Table~\ref{NPOI_table_ab} lists the results for the separation and 
position angle of $\sigma$ Ori Aa,Ab derived from the NPOI data, the semi-axes 
and position angle of the uncertainty ellipses, and the residuals compared
with the orbit fit.

\section{CTIO Spectroscopy}

We obtained new spectrocopic radial velocity measurements of $\sigma$ Orionis 
Aa,Ab using the 1.5\,m telescope at CTIO.  We obtained 40 observations on 
29 nights using the Fiber Echelle (FE) Spectrograph\footnote{\url
http://www.ctio.noao.edu/$\sim$atokovin/echelle/FECH-overview.html}
($R=25,000$, $\lambda$ =  4800--7000 \AA) between UT 2008 September 23 
and 2009 February 21.  Additional observations were obtained using the 
Chiron fiber-fed echelle spectrometer \citep{tokovinin13} equipped
with an image slicer ($R=78,000$, $\lambda$ = 4550--8800 \AA) on 10 nights
between UT 2012 November 4 and 2013 February 2 and
11 nights between UT 2016 January 21 and March 27.
The Chiron observations were concentrated near periastron passage
of the close pair.

All of the spectra were corrected to a heliocentric velocity scale prior to
measurement.  For the FE data, we measured the velocities of the He I
5876 line because it is in the same order as the interstellar Na I D
lines, which provide a good velocity fiducial. For the Chiron data, we fit
five He I lines ($\lambda\lambda$ 4713, 4921, 5876, 6678, and 7065 \AA) 
and He II ($\lambda=4686$ \AA).  The He I lines are stronger in the cooler, 
less massive component while He II is stronger in the more rapidly rotating 
hotter star.  We fit two Gaussian components to each line to measure the radial 
velocities of both components.  We allowed the central wavelength, width, and
amplitude of the Gaussian components to vary independently for each fit.  The 
He II 4686 and He I 6678 line profiles are fairly clean, while contamination 
from weak lines from the cooler star in the three bluest He I lines required 
fitting up to three additional Gaussian components.  We treat these additional
components as nuisance parameters.  Telluric lines are present at 5876 \AA\ and
are a significant problem at 7065 \AA.  We generated a telluric spectrum 
by filtering these spectra with a low-pass filter to remove the higher frequency 
narrow lines while preserving the He I line profiles, and then fit these 
``cleaned" spectra.

To check the wavelength stability, we measured the insterstellar Na D1 and D2
lines in all of the spectra.  At the lower resolution of the FE, contamination 
by the telluric lines can distort the Na D profiles as they shift due to the
 heliocentric correction.  In fact, there is a small annual distortion in the
measured velocity of the Na D lines in the FE spectra.  The median radial 
velocities measured from the FE spectra are $21.72 \pm 0.41$ km\,s$^{-1}$ for 
Na D1 and $22.63 \pm 0.69$ km\,s$^{-1}$ for Na I D2, where the uncertainties are 
the standard deviations from the mean.
With the higher resolution Chiron spectra we were able to fit both
interstellar components (a weaker line at about $+10$ km\,s$^{-1}$) and avoid
the stronger telluric features.  The stronger lines have stable
radial velocities with a median of $22.55 \pm 0.21$ km\,s$^{-1}$ for Na D1 and
$22.54 \pm 0.19$ km\,s$^{-1}$ for Na D2.  The Chiron instrumental resolution
is about 3.8 km\,s$^{-1}$.  
There seem to be no significant offsets between the two instrument
zero-points.  \citet{hobbs69} resolved the Na D
lines into two components with velocities of 20.5 and 24.0 km\,s$^{-1}$ at
higher spectral resolution ($\sim$ 0.51 km\,s$^{-1}$ ); these 
would average to 22.3 km\,s$^{-1}$, consistent with our measurements.

The median radial velocities of $\sigma$ Ori Aa and Ab, measured from the 
selected spectral lines, are presented in Table~\ref{tab.rv}.  Based on the 
Gaussian line widths, we derived rotational velocities of 
$v \sin{i} \approx 125$ km\,s$^{-1}$ for Aa and 
$v \sin{i} \approx 43$ km\,s$^{-1}$ for Ab (assuming no limb-darkening). 
We did not fit for the weak and broad stationary lines from
$\sigma$~Ori~B, which are difficult to detect without detailed modeling
\citep{simondiaz11,simondiaz15}.  Because the spectral profiles of
$\sigma$~Ori~B are so shallow, their presence creates only a slight depression
of the continuum near line center and has little influence on the velocity
measurements of components Aa and Ab.  

\begin{figure*}
  \plottwo{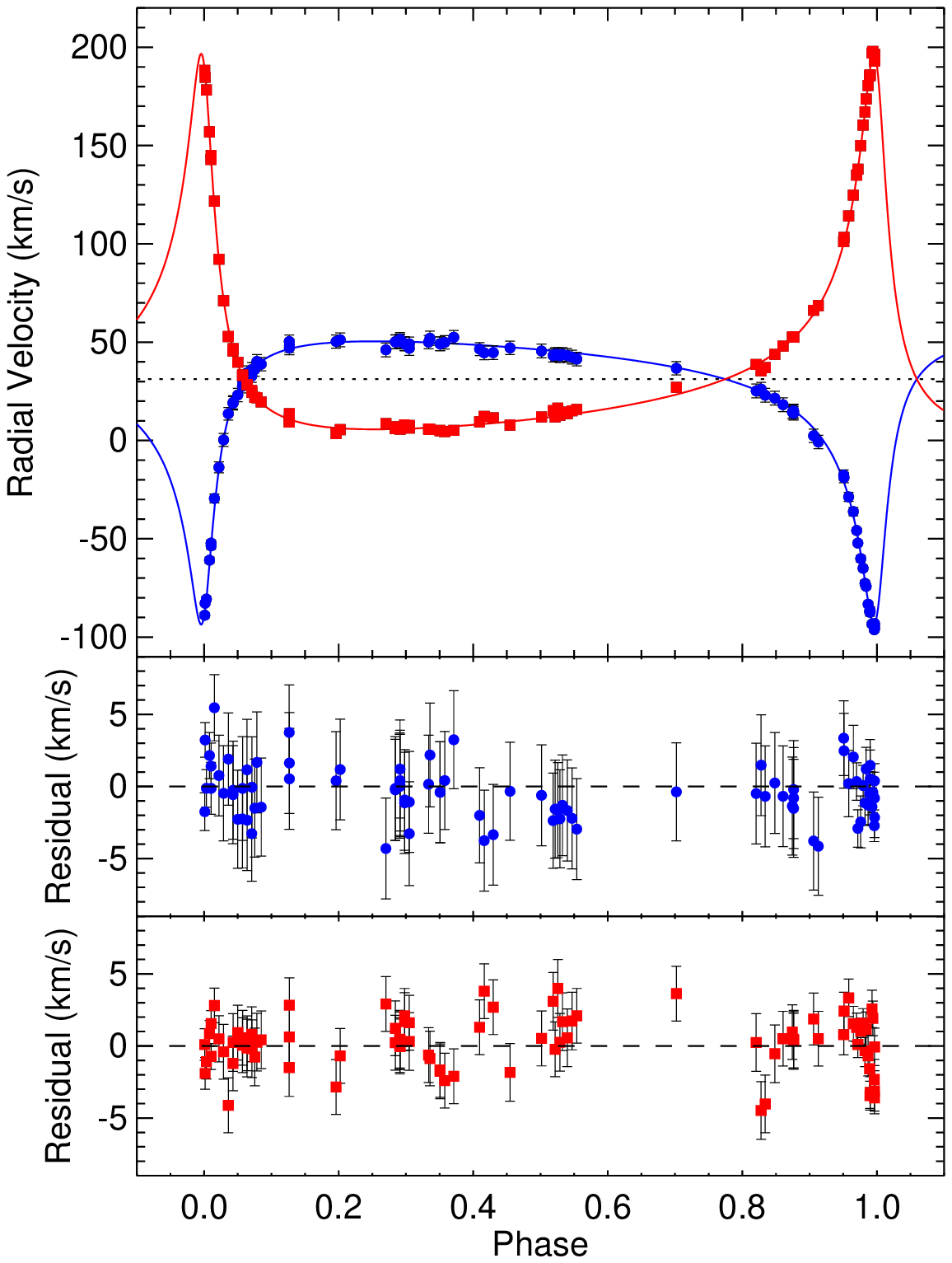}{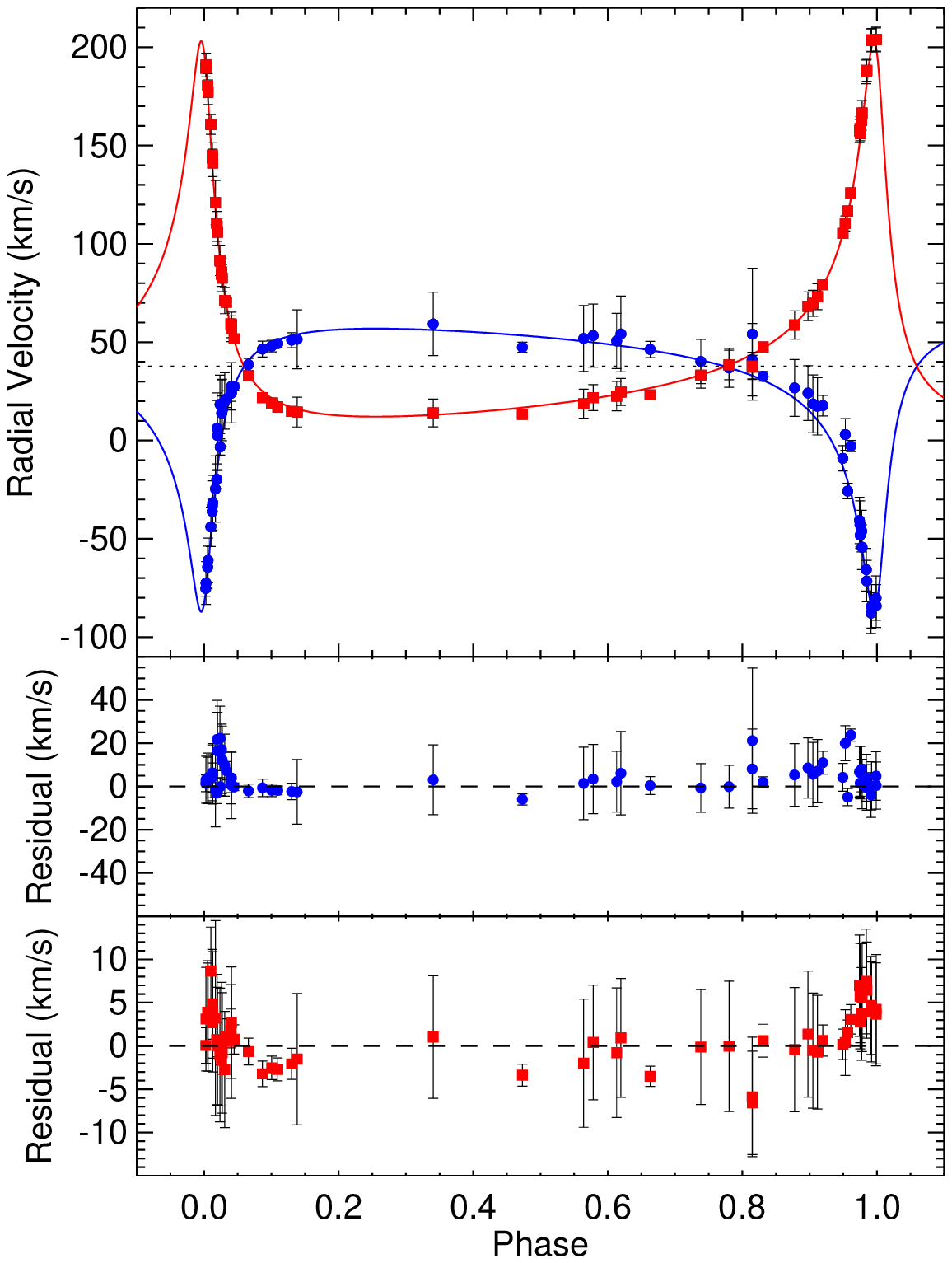}
  \caption{Radial velocities of $\sigma$ Orionis Aa,Ab published by \citet[][left]{simondiaz15} and measured at CTIO (right).  The blue circles show the velocities measured for Aa and the red squares show the velocities for Ab.  Overplotted are the radial velocity curves derived from the simultaneous orbit fit to both sets of spectroscopic data and the interferometric positions measured at the CHARA Array.  The residuals for each component are shown in the lower panels.}
\label{fig.orbit_sb2}
\end{figure*}

The first two columns of Table~\ref{tab.orb} show the spectroscopic
orbital parameters derived by \citet{simondiaz15} compared with those
derived from the CTIO radial velocities.  There are systematic differences
between the radial velocity semi-amplitudes ($K_{Aa}$, $K_{Ab}$) and the systemic 
velocity $\gamma$ derived from each set of data.  
Sim\'{o}n-D\'{i}az et al.\ cross-correlated the spectra against atmospheric
models, fitting many lines simultaneously, which could account 
for the higher precision of their velocity semi-amplitudes.  
The systematic differences could result from the different methods used to fit 
the blended lines, as well as differences in the wavelength calibration.  
A comparison of the radial velocity measurements is shown in Figure~\ref{fig.orbit_sb2}.  
A simultaneous orbit fit to both sets of data, along with the interferometric 
positions, is discussed in Section~\ref{sect.orbAaAb}.

\section{Orbits and Derived Properties of the $\sigma$ Orionis Triple}

\subsection{Visual and Spectroscopic Orbit of the Close Pair $\sigma$ Orionis Aa,Ab}
\label{sect.orbAaAb}

\begin{figure}
  \plotone{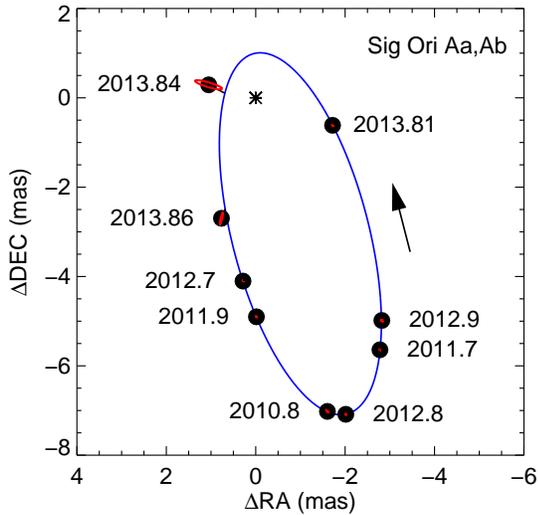}
  \caption{Visual orbit of $\sigma$ Orionis Aa,Ab based on the simultaneous fit to the interferometric positions measured using the MIRC beam combiner at the CHARA Array, the radial velocities published by \citet{simondiaz15}, and the CTIO radial velocities.  The black circles mark the position of the companion Ab relative to Aa while the red ellipses show the size and orientation of the 1\,$\sigma$ uncertainties.  The arrow indicates the direction of motion.}
\label{fig.orbit_vb}
\end{figure}

We fit a simultaneous orbit to the higher precision interferometric positions of $\sigma$ Orionis Aa,Ab measured with CHARA in Table~\ref{tab.sepPA_mirc}, the published radial velocities reported by \citet{simondiaz15}, and the CTIO radial velocities in Table~\ref{tab.rv}.  We compare the fit to the NPOI positions of the close pair in Section~\ref{sect.npoi_orb}.  Before computing the joint orbit fit, we fit each set of data independently and scaled the measurement uncertainties to force the reduced $\chi^2_\nu = 1$ for each of the CHARA and two radial velocity sets.  The measurement uncertainties in the interferometric positions were increased by a factor of 2.24, indicating that the error bars from the covariance matrix are underestimated; we report the scaled uncertainties in Table~\ref{tab.sepPA_mirc}.  The reduced $\chi^2_\nu$ for the radial velocity data from \citet{simondiaz15} was already close to 1, so we did not adjust those uncertainties.  The measurement errors for the CTIO radial velocities were decreased by a factor of 0.66 (the uncertainties listed in Table~\ref{tab.rv} are the unscaled values).  Using the scaled uncertainties, we then fit the measured positions and radial velocities simultaneously using a Newton-Raphson method to minimize $\chi^2$ by calculating a first-order Taylor expansion for the equations of orbital motion.  The last column of Table~\ref{tab.orb} provides the orbital parameters determined from the joint fit, including the period $P$, time of periastron passage $T$, eccentricity $e$, angular semi-major axis $a$, inclination $i$, position angle of the line of nodes $\Omega$, argument of periastron passage for the primary $\omega_{\rm Aa}$, and the radial velocity amplitudes of the primary and secondary $K_{Aa}$ and $K_{Ab}$.  We allowed for a shift in the systemic velocity $\gamma$ between the two sets of spectroscopic radial velocities.  Figures \ref{fig.orbit_sb2} and \ref{fig.orbit_vb} show the simultaneous spectroscopic and visual orbit fits.  The orbital phase and radial velocity residuals for the simultaneous fit are listed in Table~\ref{tab.rv}.  For comparison, we also list in Table~\ref{tab.orb} the orbital parameters determined from the fits to each set of data independently.  The velocity amplitudes derived from the joint fit depend more on the higher precision radial velocities published by \citet{simondiaz15} than on the CTIO radial velocities.

The uncertainty in the wavelength calibration of $\pm$0.25\% for MIRC \citep{monnier12}, will systematically increase or decrease the angular separations measured for the close pair.  To account for this, we varied all of the separations systematically by $\pm$0.25\% and re-fit the orbital parameters.  The second uncertainty listed for the semi-major axis for the simultaneous fit in Table~\ref{tab.orb} shows the size of the systematic uncertainty on the orbital fit.

\subsection{Visual Orbit of the Wide Pair $\sigma$ Orionis A,B}
\label{sect.npoi_orb}

\begin{figure}
\plotone{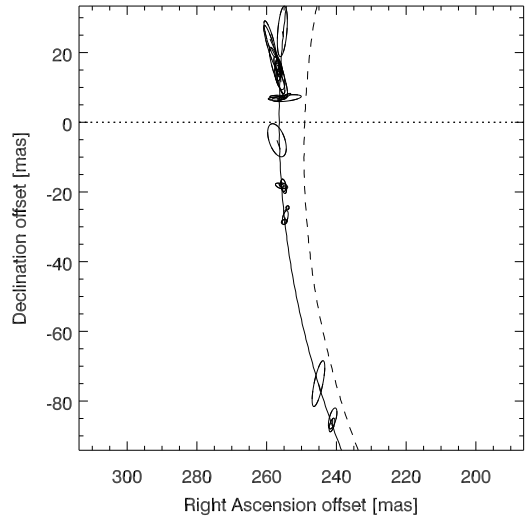}
\caption{Orbital motion of the tertiary component ($\sigma$ Ori B) 
relative to the photo-center of the close pair ($\sigma$ Ori Aa,Ab).  The
error ellipses show the positions measured with NPOI in 
Table~\ref{NPOI_table_abc}. The solid line is the best-fit orbit and the 
dashed line is the orbit computed by \citet{turner08}.}
\label{NPOI_fig_orbit_abc}
\end{figure}

\begin{figure}
\plotone{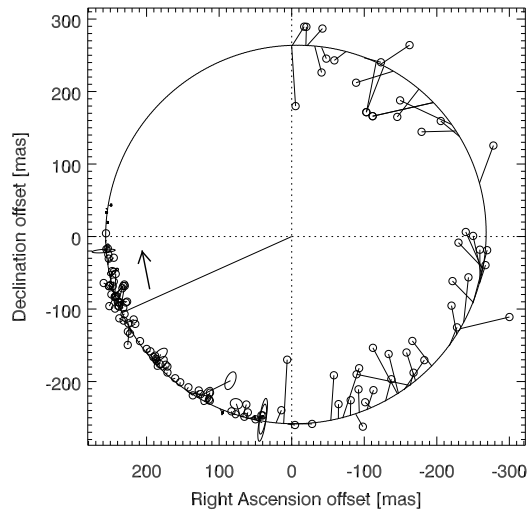}
\caption{Orbit of the tertiary ($\sigma$ Ori B) shown with all of the 
measurements available in the Washington Double Star Catalog.  The high
precision measurements in the north-east (upper-left) quadrant are the AstraLux
measurements published by \citet{simondiaz15}.  The solid line indicates 
periastron and the arrow shows the direction of motion.
}
\label{NPOI_fig_orbit_wds}
\end{figure}

As discussed in Section~\ref{sect.npoi_astrom}, we computed the orbital elements for the tertiary 
orbit ($\sigma$ Ori A,B) based on the positions derived from the NPOI data 
(Table~\ref{NPOI_table_abc}) together with all available measurements from the 
Washington Double Star Catalog.  The orbital elements are given in 
Table~\ref{NPOI_table_orbit}.  The tertiary orbit is shown with
the NPOI measurements in Figure~\ref{NPOI_fig_orbit_abc} and all
available measurements in Figure~\ref{NPOI_fig_orbit_wds}.
We used the Levenberg-Marquardt method for fitting the orbital elements to the data.

\begin{figure}
\plotone{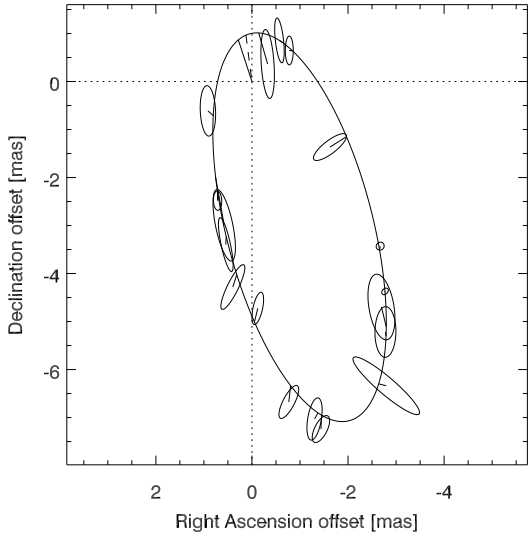}
\caption{Orbital positions of $\sigma$ Ori Ab relative to Aa as measured from the NPOI observations.  The single VLTI AMBER observation is included as well ($\Delta\alpha=0.5$ mas, $\Delta\delta=-4.5$ mas).  Overplotted is the orbit determined from the analysis of the CHARA MIRC observations.
}
\label{NPOI_fig_orbit_ab}
\end{figure}

In addition to measuring the positions of the secondary and tertiary components during each individual night, we also fit the orbital parameters directly to the NPOI visibility data.  This has the advantage of better constraining the system parameters which do not change from night to night.  We used the Levenberg-Marquardt procedure \citep{press92} to perform a non-linear least-squares fit to the visibility data and solved simultaneously for the orbital parameters for both orbits (Aa,Ab and A,B) and the magnitude difference between each component.  We fixed the component diameters at values of 0.27 mas, 0.21 mas, and 0.17 mas for components Aa, Ab, and B, respectively.  These diameters were estimated based on their $V$ magnitudes (derived from the fitted magnitude differences and the total magnitude of the system $V=3.80$ mag) and adopting the same $(V-K)$ color of $-0.69$ for all three components (as they are all of early type; derived using the total magnitude of the system of $K=4.49$ mag).  Such small diameters are unresolved on the baselines used during our observations.  Because of the large number of fit parameters, the numerical partial derivatives of $\chi^2$ with respect to the model parameters were based on step sizes optimized to give similar increases in $\chi^2$ for each parameter.  The reduced $\chi^2$ of the fits to the visibility data was 1.66 ($\chi^2=3.1$ without the floating calibration).  The orbital elements of $\sigma$ Orionis Aa,Ab derived from the NPOI data agree with the parameters derived from the CHARA data within 0.1$-$2.0\,$\sigma$, but are less precise so we do not report the NPOI parameters explicitly.  However, Figure~\ref{NPOI_fig_orbit_ab} shows that the NPOI astrometric positions are in good agreement with the CHARA orbit.

\begin{figure}
\plotone{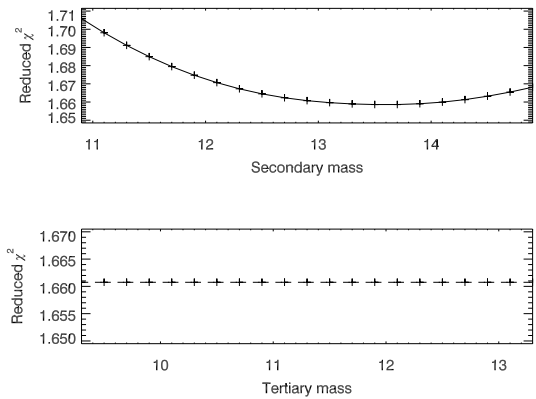}
\caption{Reduced $\chi^2$ as a function of secondary mass (top panel; $\sigma$ Ori Ab)
and tertiary mass (bottom panel; $\sigma$ Ori B).
}
\label{NPOI_fig_secmass}
\end{figure}

When the tertiary is detected by the NPOI, it can be used as a
phase reference to measure the absolute motions of component Aa and
Ab relative to their center of mass.  This provides an independent estimate 
of their mass ratio $M_{\rm Ab}/M_{\rm Aa}$.  In Fig.~\ref{NPOI_fig_secmass}
we show the reduced $\chi^2$ of the fit to the NPOI visibility data as
a function of the mass of Ab, which shows a minimum at $M_{\rm Ab} = 13.5 \pm 0.4$
$M_\odot$, assuming a fixed mass for the primary of the $M_{\rm Aa}$ = 16.9 $M_\odot$. 
Away from this value, the relative positions of the three components
change as the center of mass of the close binary shifts because of the
change in the mass ratio between Aa and Ab.  As a check,
we also show that the $\chi^2$ does not vary with tertiary mass, as may
be expected from the fact that this changes only the phase center of the
triple system. Given the uncertainty of $M_{\rm Ab}$ fit to the NPOI data, 
we do not consider the difference from the value of the dynamical mass 
derived in Section~\ref{sect.mass} to be significant.

\subsection{Stellar Masses and Distance}
\label{sect.mass}

Using the orbital parameters of the close pair $\sigma$ Ori Aa,Ab in Table~\ref{tab.orb}, we derive dynamical masses of $M_{\rm Aa}$ = 16.99  $\pm$ 0.20 $M_\odot$ and $M_{\rm Ab}$ = 12.81  $\pm$ 0.18 $M_\odot$.  The orbital parallax of $\pi$ = 2.5806 $\pm$ 0.0088 mas gives a distance $d =$  387.5 $\pm$ 1.3 pc to the $\sigma$ Orionis system.  The total mass contained in the triple can be derived from the orbital parameters of the wide pair $\sigma$ Ori A,B in Table~\ref{NPOI_table_orbit} and the orbital parallax $\pi$,
\begin{eqnarray}
M_{\rm tot} &=& M_{\rm Aa} + M_{\rm Ab} + M_{\rm B} = a_{\rm AB}^3/(\pi^3 P^2_{\rm AB}) \nonumber \\
          &=& 41.4 \pm 1.1~M_\odot. \nonumber
\end{eqnarray}
Combined with the individual masses of Aa and Ab, this yields the mass of the tertiary component of $M_{\rm B}$ = 11.5 $\pm$ 1.2 $M_\odot$.  The derived physical properties of the $\sigma$ Orionis system are summarized in Table~\ref{tab.properties}.

\section{Discussion}

\subsection{Comparison of stellar masses with evolutionary models}

\citet{simondiaz15} compared spectroscopically derived physical properties of $\sigma$ Ori Aa, Ab, and B with evolutionary tracks for rotating stars in the Milky Way computed by \citet{brott11} to derive evolutionary masses of $M_{\rm Aa}$ = 20.0 $\pm$ 1.0 $M_\odot$, $M_{\rm Ab}$ = 14.6 $\pm$ 0.8 $M_\odot$, and $M_{\rm B}$ = 13.6 $\pm$ 1.1 $M_\odot$.  These masses are systematically larger than the dynamical masses we computed in Table~\ref{tab.properties}.  Additionally, the ages derived by \citet[][Aa: 0.3$^{+1.0}_{-0.3}$ Myr, Ab: 0.9$^{+1.5}_{-0.9}$ Myr, B: 1.5$^{+1.6}_{-1.9}$ Myr]{simondiaz15} are smaller than the typical age adopted for the $\sigma$ Orionis cluster of 2$-$3 Myr \citep[e.g.,][]{sherry08}.  Future progress on resolving the discrepancies in the masses and ages could involve refining the component temperatures and luminosities, or adjusting the input parameters for the evolutionary models, especially because the evolution of massive stars is strongly dependent on their rotation and metallicity \citep{brott11,ekstrom12}.

\citet{weidner10a} studied the empirical correlation between the mass of a cluster and its most massive member.  Our dynamical mass for $\sigma$ Ori Aa of 16.99  $\pm$ 0.20 $M_\odot$ provides an additional high precision mass measurement of the most massive member of the $\sigma$ Orionis cluster.  Estimates of the total mass of the cluster range from 225 $\pm$ 30 $M_\odot$ \citep{sherry04} down to $\sim$ 150 $M_\odot$ \citep{caballero07}; these estimates are strongly dependent on the membership selection, assumed reddening, multiplicity, and evolutionary models used to estimate the masses.

\subsection{Distance to the $\sigma$ Orionis Cluster}

The distance to the $\sigma$ Orionis cluster has remained a large source of uncertainty in determining the age of the cluster and characterizing the physical properties and disk life-times for the stars, brown dwarfs, and planetary-mass members in the region.  The {\it Hipparcos} parallax of $\sigma$ Orionis itself (2.84 $\pm$ 0.91 mas) yields a distance with a large uncertainty of $352^{+166}_{-85}$ pc \citep{perryman97}.  The new reduction of the {\it Hipparcos} data gives a parallax of 3.04 $\pm$ 8.92 mas \citep{vanleeuwen07b,vanleeuwen07}, resulting in a slightly smaller distance of 329 pc, but with a much larger uncertainty. $\sigma$ Orionis presents a difficult problem for the {\it Hipparcos} analysis as it is bright and occasionally saturated, and the signals from the  three components are mixed. This required an individual component solution in the original reduction \citep{esa97}. Such individual attention was not possible in all cases for the new reduction and accounts for the large uncertainty (F.\ van Leeuwen, 2016, priv. comm.). Nevertheless, the original and new {\it Hipparcos} reductions yield parallaxes that agree within the uncertainty of the original reduction.  The orbital parallax that we measure is two orders of magnitude more precise and provides an independent check of the {\it Hipparcos} parallax for this triple star system and could be of use as a check of {\it GAIA} parallaxes for multiple stars.

Several other methods have been used to estimate the distance to $\sigma$ Orionis. \citet{francis12} computed a distance to the $\sigma$ Orionis cluster of $446 \pm 30$ pc based on the average {\it Hipparcos} parallaxes measured for 15 members.  By comparing the apparent magnitudes and the dynamical mass from the visual orbit of $\sigma$ Ori A,B with evolutionary models, \citet{caballero08d} derived a smaller distance of $334^{+25}_{-22}$ pc, or 385 pc if the system is treated as a triple.  Using main-sequence fitting to the bright members in the $\sigma$ Orionis cluster, \citet{sherry08} derived a distance of $420 \pm 30$ pc while \cite{mayne08} derived a distance of $389^{+34}_{-24}$ pc.  The variation in these distance estimates is large, although, for the most part, the values overlap within the range of their 1\,$\sigma$ uncertainties.  Our distance from the orbital parallax of the $\sigma$ Orionis multiple system of 387.5 $\pm$ 1.3 pc provides a significant improvement in the precision compared with the previous estimates of the cluster distance, and will reduce the uncertainties in future estimates of the age of the cluster based on isochrone fits.

\subsection{Alignment of the Inner and Outer Orbits}

The alignment of the orbits between the inner and outer pairs in heirarchical multiple systems can probe the initial conditions of star formation \citep{fekel81,sterzik02}.  The relative inclination between the inner and outer orbits is given by
\begin{eqnarray}
\cos{\Phi} &=& \cos{i_{\rm wide}}\cos{i_{\rm close}} \pm \sin{i_{\rm wide}}\sin{i_{\rm close}} \nonumber \\
           & & \times \cos{(\Omega_{\rm wide} - \Omega_{\rm close})}
\end{eqnarray}
\citep{fekel81} where $i_{\rm close}$ and $\Omega_{\rm close}$ are the inclination and position angle of the ascending node for the close orbit while $i_{\rm wide}$ and $\Omega_{\rm wide}$ are the same parameters for the wide orbit.  Coplanar orbits will have a relative alignment close to $\Phi = 0$.  For the wide visual pair, $\sigma$ Ori A,B, there exists a 180$^\circ$ ambiguity between $\Omega$ and $\omega$.  For the close pair, $\sigma$ Ori Aa,Ab, $\omega$ is defined by the spectroscopic orbit, so there is no ambiguity with $\Omega_{\rm close}$.  Using the orbital parameters in Tables~\ref{tab.orb} and \ref{NPOI_table_orbit}, and accounting for the ambiguity in $\Omega_{\rm wide}$, this leads to two possibilities for the relative inclination between the inner and outer orbits of $120\fdg0 \pm 2\fdg6$ or $126\fdg6 \pm 2\fdg0$.  Therefore, the alignment of the two orbits in the $\sigma$ Orionis triple are within $\sim 30^\circ$ of orthogonal.  The orbital motion of the inner pair is prograde \citep[in the direction of increasing position angles;][]{heintz78} while the motion of the outer pair is retrograde, as indicated by the directional arrows in Figures~\ref{fig.orbit_vb} and \ref{NPOI_fig_orbit_wds}.  This situation is not necessarily rare; the inner and outer orbits in the Algol triple are also nearly orthogonal, with opposing directions of motion \citep{zavala10,baron12}.

\section{Conclusions}

We obtained interferometric observations of the triple star $\sigma$ Orionis using the CHARA Array, NPOI, and VLTI.  We revised the orbital parameters for the wide A,B pair and present the first visual orbit for the close Aa,Ab pair, fit simultaneously with new and previously published radial velocities.  The orbit of the close pair is eccentric ($e \sim 0.78$) but the stars are reliably separated at periastron ($\rho_{\rm min} \sim 0.91$ mas).   Through our analysis of the orbital motion in the triple system, we derived dynamical masses of $M_{\rm Aa}$ = 16.99  $\pm$ 0.20 $M_\odot$, $M_{\rm Ab}$ = 12.81  $\pm$ 0.18 $M_\odot$, and $M_{\rm B}$ = 11.5 $\pm$ 1.2 $M_\odot$, and a distance of 387.5 $\pm$ 1.3 pc.  The orbital parallax places the $\sigma$~Orionis system about $7\%$ closer to the Sun than the Orion Nebula Cluster, which lies at a distance of $415 \pm 5$ pc based on VLBI parallaxes \citep{reid14}.

Two other bright members of the Orion OB1b association are also known triples,
$\zeta$~Ori \citep{hummel13} and $\delta$~Ori \citep{richardson15}.
The outer tertiary star appears to be a rapid rotator in each of
$\sigma$~Ori \citep[$V\sin i = 250$ km~s$^{-1}$;][]{simondiaz15},
$\zeta$~Ori  \citep[$V\sin i = 350$ km~s$^{-1}$;][]{hummel13}, and
$\delta$~Ori \citep[$V\sin i = 252$ km~s$^{-1}$;][]{richardson15}.
This suggests that the angular momentum of the natal cloud was
transformed mainly into orbital angular momentum for the stars of
the inner binary and into spin angular momentum for the outer tertiary star.
It is also possible that these triples began life as trapezium systems
of four stars in which dynamical processes led to a merger of one pair
that we see today as the distant rapid rotator.  Joint interferometric
and spectroscopic studies offer the means to determine the outcome
products of the dynamical processes of massive star formation.

\acknowledgements

We thank Deane Peterson for initially proposing to observe $\sigma$ Orionis 
with NPOI, and acknowledge his and Tom Bolton's support of the project during 
the initial phase.
We appreciate P.\ J.\ Goldfinger, Nic Scott, and Norm Vargas for 
providing operational support during the CHARA observations.
We are grateful to Ming Zhao for collecting an early set of CHARA data on 
$\sigma$~Orionis before the photometric channels were installed in MIRC.
We thank Jim Benson and the NPOI observational support staff whose efforts 
made the observations possible.
We appreciate Floor van Leeuwen for a helpful discussion on the parallaxes
of multiple stars observed by the {\it Hipparcos} mission.
We thank the referee for providing feedback to improve the manuscript.
This work is based on observations obtained with the
Georgia State University Center for High Angular
Resolution Astronomy Array at Mount Wilson Observatory.
The CHARA Array is supported by the National Science
Foundation (NSF) under Grant No. AST-1211929.  GHS and DRG 
acknowledge support from NSF Grant AST-1411654.
Institutional support has been provided from the GSU
College of Arts and Sciences and the GSU Office of the
Vice President for Research and Economic Development.
The Navy Precision Optical Interferometer is a joint project of the
Naval Research Laboratory and the US Naval Observatory, in cooperation
with Lowell Observatory and is funded by the Office of Naval Research
and the Oceanographer of the Navy.
FMW thanks Dennis Assanis, Provost of Stony Brook University, for enabling
access to Chiron spectrograph, operated by the SMARTS consortium,
through a Research Support grant.
SK acknowledges support from a European Research Council Starting Grant
(Grant Agreement No.\ 639889).
This research has made use of the SIMBAD astronomical literature database,
operated at CDS, Strasbourg, France and the Washington Double Star Catalog 
maintained at the U.S. Naval Observatory.

\facilities{CHARA, CTIO:1.5m, NPOI, VLTI}

\bibliography{ms}

\begin{thebibliography}{}
\expandafter\ifx\csname natexlab\endcsname\relax\def\natexlab#1{#1}\fi

\bibitem[{{Aldoretta} {et~al.}(2015){Aldoretta}, {Caballero-Nieves}, {Gies},
  {Nelan}, {Wallace}, {Hartkopf}, {Henry}, {Jao}, {Ma{\'{\i}}z Apell{\'a}niz},
  {Mason}, {Moffat}, {Norris}, {Richardson}, \& {Williams}}]{aldoretta15}
{Aldoretta}, E.~J., {Caballero-Nieves}, S.~M., {Gies}, D.~R., {et~al.} 2015,
  \aj, 149, 26

\bibitem[{{Aller} {et~al.}(1982){Aller}, {Appenzeller}, {Baschek}, {Duerbeck},
  {Herczeg}, {Lamla}, {Meyer-Hofmeister}, {Schmidt-Kaler}, {Scholz},
  {Seggewiss}, {Seitter}, \& {Weidemann}}]{aller82}
{Aller}, L.~H., {Appenzeller}, I., {Baschek}, B., {et~al.}, eds. 1982,
  {Landolt-B{\"o}rnstein: Numerical Data and Functional Relationships in
  Science and Technology, Vol. 2} (New York: Springer)

\bibitem[{{Armstrong} {et~al.}(1998){Armstrong}, {Mozurkewich}, {Rickard},
  {Hutter}, {Benson}, {Bowers}, {Elias}, {Hummel}, {Johnston}, {Buscher},
  {Clark}, {Ha}, {Ling}, {White}, \& {Simon}}]{armstrong98}
{Armstrong}, J.~T., {Mozurkewich}, D., {Rickard}, L.~J., {et~al.} 1998, \apj,
  496, 550

\bibitem[{{Baines} {et~al.}(2008){Baines}, {McAlister}, {ten Brummelaar},
  {Turner}, {Sturmann}, {Sturmann}, {Goldfinger}, \& {Ridgway}}]{baines08}
{Baines}, E.~K., {McAlister}, H.~A., {ten Brummelaar}, T.~A., {et~al.} 2008,
  \apj, 680, 728

\bibitem[{{Baron} {et~al.}(2012){Baron}, {Monnier}, {Pedretti}, {Zhao},
  {Schaefer}, {Parks}, {Che}, {Thureau}, {ten Brummelaar}, {McAlister},
  {Ridgway}, {Farrington}, {Sturmann}, {Sturmann}, \& {Turner}}]{baron12}
{Baron}, F., {Monnier}, J.~D., {Pedretti}, E., {et~al.} 2012, \apj, 752, 20

\bibitem[{{B{\'e}jar} {et~al.}(1999){B{\'e}jar}, {Osorio}, \&
  {Rebolo}}]{bejar99}
{B{\'e}jar}, V.~J.~S., {Osorio}, M.~R.~Z., \& {Rebolo}, R. 1999, \apj, 521, 671

\bibitem[{{B{\'e}jar} {et~al.}(2011){B{\'e}jar}, {Zapatero Osorio}, {Rebolo},
  {Caballero}, {Barrado}, {Mart{\'{\i}}n}, {Mundt}, \&
  {Bailer-Jones}}]{bejar11}
{B{\'e}jar}, V.~J.~S., {Zapatero Osorio}, M.~R., {Rebolo}, R., {et~al.} 2011,
  \apj, 743, 64

\bibitem[{{Benson} {et~al.}(2003){Benson}, {Hummel}, \&
  {Mozurkewich}}]{benson03}
{Benson}, J.~A., {Hummel}, C.~A., \& {Mozurkewich}, D. 2003, Proc. SPIE, 4838,
  358

\bibitem[{{Boden}(2000)}]{boden00}
{Boden}, A.~F. 2000, in Principles of Long Baseline Stellar Interferometry, ed.
  P.~R. {Lawson} (Pasadena, CA: JPL), 9

\bibitem[{{Bolton}(1974)}]{bolton74}
{Bolton}, C.~T. 1974, \apjl, 192, L7

\bibitem[{{Bonneau} {et~al.}(2011){Bonneau}, {Delfosse}, {Mourard}, {Lafrasse},
  {Mella}, {Cetre}, {Clausse}, \& {Zins}}]{bonneau11}
{Bonneau}, D., {Delfosse}, X., {Mourard}, D., {et~al.} 2011, \aap, 535, A53

\bibitem[{{Bonneau} {et~al.}(2006){Bonneau}, {Clausse}, {Delfosse}, {Mourard},
  {Cetre}, {Chelli}, {Cruzal{\`e}bes}, {Duvert}, \& {Zins}}]{bonneau06}
{Bonneau}, D., {Clausse}, J.-M., {Delfosse}, X., {et~al.} 2006, \aap, 456, 789

\bibitem[{{Brott} {et~al.}(2011){Brott}, {de Mink}, {Cantiello}, {Langer}, {de
  Koter}, {Evans}, {Hunter}, {Trundle}, \& {Vink}}]{brott11}
{Brott}, I., {de Mink}, S.~E., {Cantiello}, M., {et~al.} 2011, \aap, 530, A115

\bibitem[{{Burnham}(1894)}]{burnham1894}
{Burnham}, S.~W. 1894, Publications of Lick Observatory, 2, 185

\bibitem[{{Caballero}(2007)}]{caballero07}
{Caballero}, J.~A. 2007, \aap, 466, 917

\bibitem[{{Caballero}(2008{\natexlab{a}})}]{caballero08d}
{Caballero}, J.~A. 2008{\natexlab{a}}, \mnras, 383, 750

\bibitem[{{Caballero}(2008{\natexlab{b}})}]{caballero08}
{Caballero}, J.~A. 2008{\natexlab{b}}, \aap, 478, 667

\bibitem[{{Caballero}(2014)}]{caballero14}
{Caballero}, J.~A. 2014, The Observatory, 134, 273

\bibitem[{{Che} {et~al.}(2010){Che}, {Monnier}, \& {Webster}}]{che10}
{Che}, X., {Monnier}, J.~D., \& {Webster}, S. 2010, Proc. SPIE, 7734, 77342V

\bibitem[{{Chelli} {et~al.}(2009){Chelli}, {Utrera}, \& {Duvert}}]{chelli09}
{Chelli}, A., {Utrera}, O.~H., \& {Duvert}, G. 2009, \aap, 502, 705

\bibitem[{{Drimmel} {et~al.}(2003){Drimmel}, {Cabrera-Lavers}, \&
  {L{\'o}pez-Corredoira}}]{drimmel03}
{Drimmel}, R., {Cabrera-Lavers}, A., \& {L{\'o}pez-Corredoira}, M. 2003, \aap,
  409, 205

\bibitem[{{Edwards}(1976)}]{edwards76}
{Edwards}, T.~W. 1976, \aj, 81, 245

\bibitem[{{Ekstr{\"o}m} {et~al.}(2012){Ekstr{\"o}m}, {Georgy}, {Eggenberger},
  {Meynet}, {Mowlavi}, {Wyttenbach}, {Granada}, {Decressin}, {Hirschi},
  {Frischknecht}, {Charbonnel}, \& {Maeder}}]{ekstrom12}
{Ekstr{\"o}m}, S., {Georgy}, C., {Eggenberger}, P., {et~al.} 2012, \aap, 537,
  A146

\bibitem[{{ESA}(1997)}]{esa97}
{ESA}. 1997, {The HIPPARCOS and TYCHO Catalogues} (ESA SP-1200; Noordwijk:
  {ESA})

\bibitem[{{Fekel}(1981)}]{fekel81}
{Fekel}, Jr., F.~C. 1981, \apj, 246, 879

\bibitem[{{Francis} \& {Anderson}(2012)}]{francis12}
{Francis}, C., \& {Anderson}, E. 2012, Astronomy Letters, 38, 681

\bibitem[{{Frost} \& {Adams}(1904)}]{frost1904}
{Frost}, E.~B., \& {Adams}, W.~S. 1904, \apj, 19, 151

\bibitem[{{Gallenne} {et~al.}(2015){Gallenne}, {M{\'e}rand}, {Kervella},
  {Monnier}, {Schaefer}, {Baron}, {Breitfelder}, {Le Bouquin}, {Roettenbacher},
  {Gieren}, {Pietrzy{\'n}ski}, {McAlister}, {ten Brummelaar}, {Sturmann},
  {Sturmann}, {Turner}, {Ridgway}, \& {Kraus}}]{gallenne15}
{Gallenne}, A., {M{\'e}rand}, A., {Kervella}, P., {et~al.} 2015, \aap, 579, A68

\bibitem[{{Garrison}(1967)}]{garrison67}
{Garrison}, R.~F. 1967, \pasp, 79, 433

\bibitem[{{Gies}(2003)}]{gies03}
{Gies}, D.~R. 2003, in IAU Symp., Vol. 212, A Massive Star Odyssey: From Main
  Sequence to Supernova, ed. K.~{van der Hucht}, A.~{Herrero}, \& C.~{Esteban}
  (San Francisco, CA: ASP), 91

\bibitem[{{Greenstein} \& {Wallerstein}(1958)}]{greenstein58}
{Greenstein}, J.~L., \& {Wallerstein}, G. 1958, \apj, 127, 237

\bibitem[{{Haniff}(2007)}]{haniff07}
{Haniff}, C. 2007, \nar, 51, 565

\bibitem[{{Hartkopf} {et~al.}(1996){Hartkopf}, {Mason}, \&
  {McAlister}}]{hartkopf96}
{Hartkopf}, W.~I., {Mason}, B.~D., \& {McAlister}, H.~A. 1996, \aj, 111, 370

\bibitem[{{Heintz}(1974)}]{heintz74}
{Heintz}, W.~D. 1974, \aj, 79, 397

\bibitem[{{Heintz}(1978)}]{heintz78}
{Heintz}, W.~D. 1978, {Double Stars} (Dordrect, Holland: Redidel Publishing Company)

\bibitem[{{Heintz}(1997)}]{heintz97}
{Heintz}, W.~D. 1997, \apjs, 111, 335

\bibitem[{{Hern{\'a}ndez} {et~al.}(2007){Hern{\'a}ndez}, {Hartmann}, {Megeath},
  {Gutermuth}, {Muzerolle}, {Calvet}, {Vivas}, {Brice{\~n}o}, {Allen},
  {Stauffer}, {Young}, \& {Fazio}}]{hernandez07}
{Hern{\'a}ndez}, J., {Hartmann}, L., {Megeath}, T., {et~al.} 2007, \apj, 662,
  1067

\bibitem[{{Hern{\'a}ndez} {et~al.}(2014){Hern{\'a}ndez}, {Calvet}, {Perez},
  {Brice{\~n}o}, {Olguin}, {Contreras}, {Hartmann}, {Allen}, {Espaillat}, \&
  {Hernan}}]{hernandez14}
{Hern{\'a}ndez}, J., {Calvet}, N., {Perez}, A., {et~al.} 2014, \apj, 794, 36

\bibitem[{{Hobbs}(1969)}]{hobbs69}
{Hobbs}, L.~M. 1969, \apj, 158, 461

\bibitem[{{Hummel} {et~al.}(1998){Hummel}, {Mozurkewich}, {Armstrong},
  {Hajian}, {Elias}, \& {Hutter}}]{hummel98}
{Hummel}, C.~A., {Mozurkewich}, D., {Armstrong}, J.~T., {et~al.} 1998, \aj,
  116, 2536

\bibitem[{{Hummel} {et~al.}(2013){Hummel}, {Rivinius}, {Nieva}, {Stahl}, {van
  Belle}, \& {Zavala}}]{hummel13}
{Hummel}, C.~A., {Rivinius}, T., {Nieva}, M.-F., {et~al.} 2013, \aap, 554, A52

\bibitem[{{Hummel} {et~al.}(2003){Hummel}, {Benson}, {Hutter}, {Johnston},
  {Mozurkewich}, {Armstrong}, {Hindsley}, {Gilbreath}, {Rickard}, \&
  {White}}]{hummel03}
{Hummel}, C.~A., {Benson}, J.~A., {Hutter}, D.~J., {et~al.} 2003, \aj, 125,
  2630

\bibitem[{{Koenig} {et~al.}(2015){Koenig}, {Hillenbrand}, {Padgett}, \&
  {DeFelippis}}]{koenig15}
{Koenig}, X., {Hillenbrand}, L.~A., {Padgett}, D.~L., \& {DeFelippis}, D. 2015,
  \aj, 150, 100

\bibitem[{{Kraus} {et~al.}(2005){Kraus}, {Schloerb}, {Traub}, {Carleton},
  {Lacasse}, {Pearlman}, {Monnier}, {Millan-Gabet}, {Berger}, {Haguenauer},
  {Perraut}, {Kern}, {Malbet}, \& {Labeye}}]{kraus05}
{Kraus}, S., {Schloerb}, F.~P., {Traub}, W.~A., {et~al.} 2005, \aj, 130, 246

\bibitem[{{Landstreet} \& {Borra}(1978)}]{landstreet78}
{Landstreet}, J.~D., \& {Borra}, E.~F. 1978, \apjl, 224, L5

\bibitem[{{Lawson}(2000)}]{lawson00}
{Lawson}, P.~R., ed. 2000, {Principles of Long Baseline Stellar Interferometry}
  (Pasadena, CA: JPL)

\bibitem[{{Lodieu} {et~al.}(2009){Lodieu}, {Zapatero Osorio}, {Rebolo},
  {Mart{\'{\i}}n}, \& {Hambly}}]{lodieu09}
{Lodieu}, N., {Zapatero Osorio}, M.~R., {Rebolo}, R., {Mart{\'{\i}}n}, E.~L.,
  \& {Hambly}, N.~C. 2009, \aap, 505, 1115

\bibitem[{{Luhman} {et~al.}(2008){Luhman}, {Hern{\'a}ndez}, {Downes},
  {Hartmann}, \& {Brice{\~n}o}}]{luhman08}
{Luhman}, K.~L., {Hern{\'a}ndez}, J., {Downes}, J.~J., {Hartmann}, L., \&
  {Brice{\~n}o}, C. 2008, \apj, 688, 362

\bibitem[{{Lynga}(1981)}]{lynga81}
{Lynga}, G. 1981, Astronomical Data Center Bulletin, 1, 90

\bibitem[{{Maeder}(1995)}]{maeder95}
{Maeder}, A. 1995, in ASP Conf. Ser., Vol.~83, Astrophysical Applications of
  Stellar Pulsation, ed. R.~S. {Stobie} \& P.~A. {Whitelock} (San Francisco,
  CA: ASP), 1

\bibitem[{{Markwardt}(2009)}]{markwardt09}
{Markwardt}, C.~B. 2009, in ASP Conf. Ser., Vol. 411, Astronomical Data
  Analysis Software and Systems XVIII, ed. D.~A. {Bohlender}, D.~{Durand}, \&
  P.~{Dowler} (San Francisco, CA: ASP), 251

\bibitem[{{Massey} {et~al.}(2012){Massey}, {Morrell}, {Neugent}, {Penny},
  {DeGioia-Eastwood}, \& {Gies}}]{massey12}
{Massey}, P., {Morrell}, N.~I., {Neugent}, K.~F., {et~al.} 2012, \apj, 748, 96

\bibitem[{{Mayne} \& {Naylor}(2008)}]{mayne08}
{Mayne}, N.~J., \& {Naylor}, T. 2008, \mnras, 386, 261

\bibitem[{{Miczaika}(1950)}]{miczaika50}
{Miczaika}, G.~R. 1950, \apj, 111, 443

\bibitem[{{Monnier}(2003)}]{monnier03}
{Monnier}, J.~D. 2003, Reports on Progress in Physics, 66, 789

\bibitem[{{Monnier} {et~al.}(2004){Monnier}, {Berger}, {Millan-Gabet}, \& {ten
  Brummelaar}}]{monnier04}
{Monnier}, J.~D., {Berger}, J.-P., {Millan-Gabet}, R., \& {ten Brummelaar},
  T.~A. 2004, Proc. SPIE, 5491, 1370

\bibitem[{{Monnier} {et~al.}(2006){Monnier}, {Pedretti}, {Thureau}, {Berger},
  {Millan-Gabet}, {ten Brummelaar}, {McAlister}, {Sturmann}, {Sturmann},
  {Muirhead}, {Tannirkulam}, {Webster}, \& {Zhao}}]{monnier06}
{Monnier}, J.~D., {Pedretti}, E., {Thureau}, N., {et~al.} 2006, Proc. SPIE,
  6268, 62681P

\bibitem[{{Monnier} {et~al.}(2007){Monnier}, {Zhao}, {Pedretti}, {Thureau},
  {Ireland}, {Muirhead}, {Berger}, {Millan-Gabet}, {Van Belle}, {ten
  Brummelaar}, {McAlister}, {Ridgway}, {Turner}, {Sturmann}, {Sturmann}, \&
  {Berger}}]{monnier07}
{Monnier}, J.~D., {Zhao}, M., {Pedretti}, E., {et~al.} 2007, Science, 317, 342

\bibitem[{{Monnier} {et~al.}(2012){Monnier}, {Che}, {Zhao}, {Ekstr{\"o}m},
  {Maestro}, {Aufdenberg}, {Baron}, {Georgy}, {Kraus}, {McAlister}, {Pedretti},
  {Ridgway}, {Sturmann}, {Sturmann}, {ten Brummelaar}, {Thureau}, {Turner}, \&
  {Tuthill}}]{monnier12}
{Monnier}, J.~D., {Che}, X., {Zhao}, M., {et~al.} 2012, \apjl, 761, L3

\bibitem[{{Morrell} {et~al.}(2014){Morrell}, {Massey}, {Neugent}, {Penny}, \&
  {Gies}}]{morrell14}
{Morrell}, N.~I., {Massey}, P., {Neugent}, K.~F., {Penny}, L.~R., \& {Gies},
  D.~R. 2014, \apj, 789, 139

\bibitem[{{Mourard} {et~al.}(2011){Mourard}, {B{\'e}rio}, {Perraut}, {Ligi},
  {Blazit}, {Clausse}, {Nardetto}, {Spang}, {Tallon-Bosc}, {Bonneau},
  {Chesneau}, {Delaa}, {Millour}, {Stee}, {Le Bouquin}, {ten Brummelaar},
  {Farrington}, {Goldfinger}, \& {Monnier}}]{mourard11}
{Mourard}, D., {B{\'e}rio}, P., {Perraut}, K., {et~al.} 2011, \aap, 531, A110

\bibitem[{{Mozurkewich} {et~al.}(1991){Mozurkewich}, {Johnston}, {Simon},
  {Bowers}, {Gaume}, {Hutter}, {Colavita}, {Shao}, \& {Pan}}]{mozurkewich91}
{Mozurkewich}, D., {Johnston}, K.~J., {Simon}, R.~S., {et~al.} 1991, \aj, 101,
  2207

\bibitem[{{Mozurkewich} {et~al.}(2003){Mozurkewich}, {Armstrong}, {Hindsley},
  {Quirrenbach}, {Hummel}, {Hutter}, {Johnston}, {Hajian}, {Elias}, {Buscher},
  \& {Simon}}]{mozurkewich03}
{Mozurkewich}, D., {Armstrong}, J.~T., {Hindsley}, R.~B., {et~al.} 2003, \aj,
  126, 2502

\bibitem[{{Oliveira} {et~al.}(2006){Oliveira}, {Jeffries}, {van Loon}, \&
  {Rushton}}]{oliveira06}
{Oliveira}, J.~M., {Jeffries}, R.~D., {van Loon}, J.~T., \& {Rushton}, M.~T.
  2006, \mnras, 369, 272

\bibitem[{{Pe{\~n}a Ram{\'{\i}}rez} {et~al.}(2012){Pe{\~n}a Ram{\'{\i}}rez},
  {B{\'e}jar}, {Zapatero Osorio}, {Petr-Gotzens}, \&
  {Mart{\'{\i}}n}}]{pena_ramirez12}
{Pe{\~n}a Ram{\'{\i}}rez}, K., {B{\'e}jar}, V.~J.~S., {Zapatero Osorio}, M.~R.,
  {Petr-Gotzens}, M.~G., \& {Mart{\'{\i}}n}, E.~L. 2012, \apj, 754, 30

\bibitem[{{Perryman} {et~al.}(1997){Perryman}, {Lindegren}, {Kovalevsky},
  {Hoeg}, {Bastian}, {Bernacca}, {Cr{\'e}z{\'e}}, {Donati}, {Grenon},
  {Grewing}, {van Leeuwen}, {van der Marel}, {Mignard}, {Murray}, {Le Poole},
  {Schrijver}, {Turon}, {Arenou}, {Froeschl{\'e}}, \& {Petersen}}]{perryman97}
{Perryman}, M.~A.~C., {Lindegren}, L., {Kovalevsky}, J., {et~al.} 1997, \aap,
  323, L49

\bibitem[{{Petrov} {et~al.}(2007){Petrov}, {Malbet}, {Weigelt}, \&
  {Antonelli}}]{petrov07}
{Petrov}, R.~G., {Malbet}, F., {Weigelt}, G., \& {Antonelli}, e.~a. 2007, \aap,
  464, 1

\bibitem[{{Press} {et~al.}(1992){Press}, {Teukolsky}, {Vetterling}, \&
  {Flannery}}]{press92}
{Press}, W.~H., {Teukolsky}, S.~A., {Vetterling}, W.~T., \& {Flannery}, B.~P.
  1992, {Numerical recipes in C. The art of scientific computing} (New York,
  NY: Cambridge Univ. Press)

\bibitem[{{Reid} {et~al.}(2014){Reid}, {Menten}, {Brunthaler}, {Zheng}, {Dame},
  {Xu}, {Wu}, {Zhang}, {Sanna}, {Sato}, {Hachisuka}, {Choi}, {Immer},
  {Moscadelli}, {Rygl}, \& {Bartkiewicz}}]{reid14}
{Reid}, M.~J., {Menten}, K.~M., {Brunthaler}, A., {et~al.} 2014, \apj, 783, 130

\bibitem[{{Richardson} {et~al.}(2015){Richardson}, {Moffat}, {Gull}, {Lindler},
  {Gies}, {Corcoran}, \& {Chen{\'e}}}]{richardson15}
{Richardson}, N.~D., {Moffat}, A.~F.~J., {Gull}, T.~R., {et~al.} 2015, \apj,
  808, 88

\bibitem[{{Sacco} {et~al.}(2008){Sacco}, {Franciosini}, {Randich}, \&
  {Pallavicini}}]{sacco08}
{Sacco}, G.~G., {Franciosini}, E., {Randich}, S., \& {Pallavicini}, R. 2008,
  \aap, 488, 167

\bibitem[{{Schaefer} {et~al.}(2010){Schaefer}, {Gies}, {Monnier}, {Richardson},
  {Touhami}, {Zhao}, {Che}, {Pedretti}, {Thureau}, {ten Brummelaar},
  {McAlister}, {Ridgway}, {Sturmann}, {Sturmann}, {Turner}, {Farrington}, \&
  {Goldfinger}}]{schaefer10}
{Schaefer}, G.~H., {Gies}, D.~R., {Monnier}, J.~D., {et~al.} 2010, \aj, 140,
  1838

\bibitem[{{Sch{\"o}ller}(2007)}]{scholler07}
{Sch{\"o}ller}, M. 2007, \nar, 51, 628

\bibitem[{{Sherry} {et~al.}(2004){Sherry}, {Walter}, \& {Wolk}}]{sherry04}
{Sherry}, W.~H., {Walter}, F.~M., \& {Wolk}, S.~J. 2004, \aj, 128, 2316

\bibitem[{{Sherry} {et~al.}(2008){Sherry}, {Walter}, {Wolk}, \&
  {Adams}}]{sherry08}
{Sherry}, W.~H., {Walter}, F.~M., {Wolk}, S.~J., \& {Adams}, N.~R. 2008, \aj,
  135, 1616

\bibitem[{{Sim{\'o}n-D{\'{\i}}az} {et~al.}(2011){Sim{\'o}n-D{\'{\i}}az},
  {Caballero}, \& {Lorenzo}}]{simondiaz11}
{Sim{\'o}n-D{\'{\i}}az}, S., {Caballero}, J.~A., \& {Lorenzo}, J. 2011, \apj,
  742, 55

\bibitem[{{Sim{\'o}n-D{\'{\i}}az} {et~al.}(2015){Sim{\'o}n-D{\'{\i}}az},
  {Caballero}, {Lorenzo}, {Ma{\'{\i}}z Apell{\'a}niz}, {Schneider},
  {Negueruela}, {Barb{\'a}}, {Dorda}, {Marco}, {Montes}, {Pellerin},
  {Sanchez-Bermudez}, {S{\'o}dor}, \& {Sota}}]{simondiaz15}
{Sim{\'o}n-D{\'{\i}}az}, S., {Caballero}, J.~A., {Lorenzo}, J., {et~al.} 2015,
  \apj, 799, 169

\bibitem[{{Sterzik} \& {Tokovinin}(2002)}]{sterzik02}
{Sterzik}, M.~F., \& {Tokovinin}, A.~A. 2002, \aap, 384, 1030

\bibitem[{{Struve}(1837)}]{struve1837}
{Struve}, F.~G.~W. 1837, Astronomische Nachrichten, 14, 249

\bibitem[{{Tatulli} {et~al.}(2007){Tatulli}, {Millour}, {Chelli}, {Duvert},
  {Acke}, {Hernandez Utrera}, {Hofmann}, {Kraus}, {Malbet}, {M{\`e}ge},
  {Petrov}, {Vannier}, {Zins}, {Antonelli}, {Beckmann}, {Bresson}, {Dugu{\'e}},
  {Gennari}, {Gl{\"u}ck}, {Kern}, {Lagarde}, {Le Coarer}, {Lisi}, {Perraut},
  {Puget}, {Rantakyr{\"o}}, {Robbe-Dubois}, {Roussel}, {Weigelt}, {Accardo},
  {Agabi}, {Altariba}, {Arezki}, {Aristidi}, {Baffa}, {Behrend}, {Bl{\"o}cker},
  {Bonhomme}, {Busoni}, {Cassaing}, {Clausse}, {Colin}, {Connot},
  {Delboulb{\'e}}, {Domiciano de Souza}, {Driebe}, {Feautrier}, {Ferruzzi},
  {Forveille}, {Fossat}, {Foy}, {Fraix-Burnet}, {Gallardo}, {Giani}, {Gil},
  {Glentzlin}, {Heiden}, {Heininger}, {Kamm}, {Kiekebusch}, {Le Contel}, {Le
  Contel}, {Lesourd}, {Lopez}, {Lopez}, {Magnard}, {Marconi}, {Mars},
  {Martinot-Lagarde}, {Mathias}, {Monin}, {Mouillet}, {Mourard}, {Nussbaum},
  {Ohnaka}, {Pacheco}, {Perrier}, {Rabbia}, {Rebattu}, {Reynaud}, {Richichi},
  {Robini}, {Sacchettini}, {Schertl}, {Sch{\"o}ller}, {Solscheid}, {Spang},
  {Stee}, {Stefanini}, {Tallon}, {Tallon-Bosc}, {Tasso}, {Testi}, {Vakili},
  {von der L{\"u}he}, {Valtier}, \& {Ventura}}]{tatulli07}
{Tatulli}, E., {Millour}, F., {Chelli}, A., {et~al.} 2007, \aap, 464, 29

\bibitem[{{ten Brummelaar} {et~al.}(2005){ten Brummelaar}, {McAlister},
  {Ridgway}, {Bagnuolo}, {Turner}, {Sturmann}, {Sturmann}, {Berger}, {Ogden},
  {Cadman}, {Hartkopf}, {Hopper}, \& {Shure}}]{tenbrummelaar05}
{ten Brummelaar}, T.~A., {McAlister}, H.~A., {Ridgway}, S.~T., {et~al.} 2005,
  \apj, 628, 453

\bibitem[{{Tokovinin} {et~al.}(2013){Tokovinin}, {Fischer}, {Bonati},
  {Giguere}, {Moore}, {Schwab}, {Spronck}, \& {Szymkowiak}}]{tokovinin13}
{Tokovinin}, A., {Fischer}, D.~A., {Bonati}, M., {et~al.} 2013, \pasp, 125,
  1336

\bibitem[{{Turner} {et~al.}(2008){Turner}, {ten Brummelaar}, {Roberts},
  {Mason}, {Hartkopf}, \& {Gies}}]{turner08}
{Turner}, N.~H., {ten Brummelaar}, T.~A., {Roberts}, L.~C., {et~al.} 2008, \aj,
  136, 554

\bibitem[{{van Belle} {et~al.}(2009){van Belle}, {Creech-Eakman}, \&
  {Hart}}]{vanbelle09}
{van Belle}, G.~T., {Creech-Eakman}, M.~J., \& {Hart}, A. 2009, \mnras, 394,
  1925

\bibitem[{{van Leeuwen}(2007{\natexlab{a}})}]{vanleeuwen07b}
{van Leeuwen}, F. 2007{\natexlab{a}}, {Hipparcos, the New Reduction of the Raw
  Data} (Astrophys. Space Sci. Lib. Vol 350; Dordrecht: Springer)

\bibitem[{{van Leeuwen}(2007{\natexlab{b}})}]{vanleeuwen07}
{van Leeuwen}, F. 2007{\natexlab{b}}, \aap, 474, 653

\bibitem[{{Walter} {et~al.}(2008){Walter}, {Sherry}, {Wolk}, \&
  {Adams}}]{walter08}
{Walter}, F.~M., {Sherry}, W.~H., {Wolk}, S.~J., \& {Adams}, N.~R. 2008, in
  Handbook of Star Forming Regions, Volume I: The Northern Sky, ed.
  B.~{Reipurth} (San Francisco, CA: ASP Monograph Publications), 732

\bibitem[{{Walter} {et~al.}(1997){Walter}, {Wolk}, {Freyberg}, \&
  {Schmitt}}]{walter97}
{Walter}, F.~M., {Wolk}, S.~J., {Freyberg}, M., \& {Schmitt}, J.~H.~M.~M. 1997,
  \memsai, 68, 1081

\bibitem[{{Walter} {et~al.}(1998){Walter}, {Wolk}, \& {Sherry}}]{walter98}
{Walter}, F.~M., {Wolk}, S.~J., \& {Sherry}, W. 1998, in ASP Conf. Ser., Vol.
  154, Cool Stars, Stellar Systems, and the Sun, ed. R.~A. {Donahue} \& J.~A.
  {Bookbinder} (San Francisco, CA: ASP), 1793

\bibitem[{{Weidner} {et~al.}(2010){Weidner}, {Kroupa}, \&
  {Bonnell}}]{weidner10a}
{Weidner}, C., {Kroupa}, P., \& {Bonnell}, I.~A.~D. 2010, \mnras, 401, 275

\bibitem[{{Weidner} \& {Vink}(2010)}]{weidner10}
{Weidner}, C., \& {Vink}, J.~S. 2010, \aap, 524, A98

\bibitem[{{Zapatero Osorio} {et~al.}(2000){Zapatero Osorio}, {B{\'e}jar},
  {Mart{\'{\i}}n}, {Rebolo}, {Barrado y Navascu{\'e}s}, {Bailer-Jones}, \&
  {Mundt}}]{zapatero00}
{Zapatero Osorio}, M.~R., {B{\'e}jar}, V.~J.~S., {Mart{\'{\i}}n}, E.~L.,
  {et~al.} 2000, Science, 290, 103

\bibitem[{{Zavala} {et~al.}(2010){Zavala}, {Hummel}, {Boboltz}, {Ojha},
  {Shaffer}, {Tycner}, {Richards}, \& {Hutter}}]{zavala10}
{Zavala}, R.~T., {Hummel}, C.~A., {Boboltz}, D.~A., {et~al.} 2010, \apjl, 715,
  L44

\end{thebibliography}

\mbox

\clearpage

\onecolumngrid
\begin{deluxetable}{llllrrr} 
\tablewidth{0pt}
\tablecaption{CHARA MIRC Observation Log for $\sigma$ Orionis.  \label{tab.log}} 
\tablehead{ 
\colhead{UT Date} & \colhead{HJD $-$}   & \colhead{Configuration} & \colhead{Calibrators} & \colhead{Number of}   & \colhead{Number of}      & \colhead{Seeing} \\
\colhead{}        & \colhead{2,400,000} & \colhead{}              & \colhead{}            & \colhead{$V^2$}       & \colhead{Closure Phases}  & \colhead{$r_0$ (cm)}}
\startdata 
2010 Nov 04 &  55505.061 & S2-W1-W2          & HD 33256           &  24~~~~~~ &   8~~~~~~~ &  6.6~~   \\
2010 Nov 05 &  55506.014 & S1-W1-W2          & HD 33256           &  48~~~~~~ &  16~~~~~~~ &  7.3~~  \\
2011 Sep 29 &  55834.004 & S1-S2-E1-E2-W1-W2 & HD 25490, HD 33256 & 200~~~~~~ & 240~~~~~~~ & 12.0~~  \\
2011 Dec 09 &  55904.836 & S1-S2-E1-E2-W1-W2 & HD 25490, HD 33256 & 319~~~~~~ & 364~~~~~~~ &  6.8~~ \\
2012 Sep 15 &  56186.016 & S1-S2-E1-E2-W1-W2 & HD 33256, HD 43318 & 240~~~~~~ & 320~~~~~~~ & 13.2~~  \\
2012 Oct 31 &  56231.907 & S1-S2-E1-E2-W1-W2 & HD 33256, HD 43318 & 220~~~~~~ & 248~~~~~~~ & 15.0~~  \\
2012 Dec 09 &  56270.813 & S1-S2-E1-E2-W1-W2 & HD 25490, HD 33256, HD 43318 & 168~~~~~~ & 168~~~~~~~ &  6.1~~   \\
2013 Oct 21 &  56586.948 & S1-E1-E2-W1-W2    & HD 25490, HD 33256, HD 43318 & 205~~~~~~ & 144~~~~~~~ & 12.1~~  \\
2013 Nov 03 &  56599.926 & S1-W1-W2          & HD 33256, HD 55185 &  47~~~~~~ &  11~~~~~~~ &  7.6~~  \\
2013 Nov 11 &  56607.886 & E1-E2-W1-W2       & HD 33256, HD 43318 &  46~~~~~~ &  24~~~~~~~ & 10.7~~ \\  
\enddata 
\end{deluxetable}

\begin{deluxetable}{lcl} 
\tablewidth{0pt}
\tablecaption{Adopted Calibrator Angular Diameters for CHARA MIRC Observations} 
\tablehead{
\colhead{Calibrator} & \colhead{Angular Diameter} & \colhead{Reference} \\
\colhead{} & \colhead{(mas)} & \colhead{} }
\startdata 
HD 25490 & 0.599 $\pm$ 0.020 &  SED fit - this work \\ 
HD 33256 & 0.655 $\pm$ 0.018 &  SED fit - this work \\ 
HD 43318 & 0.491 $\pm$ 0.030 &  \citet{baines08} \\
HD 55185 & 0.474 $\pm$ 0.014 &  \citet{mourard11} 
\label{tab.cal}
\enddata 
\end{deluxetable}  

\clearpage

\onecolumngrid
\begin{deluxetable}{llrrrrrrrr} 
\tablewidth{0pt}
\tablecaption{Positions of $\sigma$ Orionis Aa,Ab measured with CHARA MIRC} 
\tablehead{ 
\colhead{UT Date} & \colhead{HJD $-$} & \colhead{$\rho$} & \colhead{$\theta$} & \colhead{$\sigma_{\rm maj}$} & \colhead{$\sigma_{\rm min}$} & \colhead{$\phi$}    & \colhead{$f_{Aa}$} & \colhead{$f_{Ab}$} & \colhead{$f_{B}$} \\
\colhead{}        & \colhead{2,400,000}    & \colhead{(mas)} & \colhead{(deg)} & \colhead{(mas)}            & \colhead{(mas)}           & \colhead{(deg)} & \colhead{}         & \colhead{}        & \colhead{} }
\startdata 
2010 Nov 05  &  55506.014  & 7.2007 & 192.867 & 0.0430 & 0.0036 &  41.16 & 0.4655 $\pm$ 0.0051 & 0.2704 $\pm$ 0.0036 & 0.2640 $\pm$ 0.0062 \\
2011 Sep 29  &  55834.004  & 6.2883 & 206.224 & 0.0046 & 0.0031 & 156.37 & 0.4347 $\pm$ 0.0029 & 0.2549 $\pm$ 0.0020 & 0.3105 $\pm$ 0.0035 \\
2011 Dec 09  &  55904.836  & 4.9017 & 180.184 & 0.0098 & 0.0076 & 132.28 & 0.4167 $\pm$ 0.0039 & 0.2477 $\pm$ 0.0037 & 0.3356 $\pm$ 0.0053 \\
2012 Sep 15  &  56186.016  & 4.1128 & 176.028 & 0.0047 & 0.0019 & 142.73 & 0.4986 $\pm$ 0.0045 & 0.2989 $\pm$ 0.0027 & 0.2025 $\pm$ 0.0053 \\
2012 Oct 31  &  56231.907  & 7.3658 & 195.874 & 0.0086 & 0.0032 & 164.21 & 0.4145 $\pm$ 0.0033 & 0.2450 $\pm$ 0.0025 & 0.3405 $\pm$ 0.0042 \\
2012 Dec 09  &  56270.813  & 5.7271 & 209.509 & 0.0167 & 0.0142 & 175.66 & 0.4947 $\pm$ 0.0058 & 0.2976 $\pm$ 0.0067 & 0.2077 $\pm$ 0.0088 \\
2013 Oct 21  &  56586.948  & 1.8253 & 250.315 & 0.0072 & 0.0050 & 134.36 & 0.4497 $\pm$ 0.0024 & 0.2666 $\pm$ 0.0020 & 0.2838 $\pm$ 0.0031 \\
2013 Nov 03  &  56599.926  & 1.0917 &  74.529 & 0.3161 & 0.0524 &  73.41 & 0.5992 $\pm$ 0.0497 & 0.1989 $\pm$ 0.0444 & 0.2019 $\pm$ 0.0666 \\
2013 Nov 11  &  56607.886  & 2.8034 & 164.128 & 0.1406 & 0.0148 & 166.21 & 0.5181 $\pm$ 0.0302 & 0.3888 $\pm$ 0.0300 & 0.0931 $\pm$ 0.0426 
\label{tab.sepPA_mirc}
\enddata 
\tablecomments{The size of the major and minor axes of the error ellipse for the binary positions have been scaled by a factor of 2.24 \\
to force $\chi^2_\nu = 1$ for an orbit fit to only the MIRC derived positions.  The second column repesents the median HJD for the \\ 
time of observation.} 
\end{deluxetable} 


\clearpage

\begin{deluxetable}{lllccl}
\tabletypesize{\scriptsize}
\tablecaption{NPOI Observation Log for $\sigma$ Orionis\label{NPOI_table_obs}}
\tablehead{
\colhead{}&
\colhead{}&
\colhead{}&
\colhead{Min. length}&
\colhead{Max. length}&
\colhead{}
\\
\colhead{UT Date}&
\colhead{Julian Year}&
\colhead{Triangles and baselines}&
\colhead{(m)}&
\colhead{(m)}&
\colhead{Calibrators (HD)}
\\
\colhead{(1)}&
\colhead{(2)}&
\colhead{(3)}&
\colhead{(4)}&
\colhead{(5)}&
\colhead{(6)}
}
\startdata
2001 Feb 21 & 2001.1408 & AC-AE-AW & 18 & 36 & 37128 \\
2001 Feb 22 & 2001.1435 & AC-AE-AW & 18 & 37 & 37128 56537 \\
2002 Feb 15 & 2002.1237 & AE-AW AW-AC W7-AC &  6 & 49 & 37128 \\
2006 Nov 08 & 2006.8529 & AC-AE-W7 & 12 & 39 & 25940 \\
2006 Nov 18 & 2006.8802 & AC-AE-W7 & 10 & 56 & 22192 5448 24760 25940 \\
2007 Feb 18 & 2007.1314 & AE-W7 AN0-W7 AW-W7 & 14 & 40 & 37128 87737 91316 \\
2007 Oct 19 & 2007.7975 & AC-AE-AN0 & 18 & 49 & 37128 \\
2007 Oct 20 & 2007.8000 & AE-AC AE-AN0 & 18 & 48 & 17573 37128 \\
2007 Nov 01 & 2007.8329 & AC-AE AC-AW & 15 & 43 & 17573 37128 \\
2007 Nov 03 & 2007.8384 & AC-AE & 17 & 35 & 17573 37128 \\
2008 Nov 16 & 2008.8761 & AC-AE-W7 & 15 & 52 & 19994 37128 21790 \\
2008 Nov 17 & 2008.8788 & AC-AE-W7 & 17 & 50 & 19994 37128 21790 \\
2010 Mar 18 & 2010.2085 & AE-AN0-AW AE-AN0-W7 & 23 & 53 & 37128 \\
2010 Mar 21 & 2010.2167 & AE-AN0-AW AE-AN0-W7 & 22 & 50 & 37128 \\
2010 Mar 25 & 2010.2277 & AE-AN0-AW AE-AN0-W7 & 20 & 46 & 37128 \\
2010 Mar 29 & 2010.2386 & AE-AN0-W7 AW-AE & 26 & 47 & 37128 \\
2011 Jan 29 & 2011.0766 & AC-AW-E6 & 17 & 52 & 24760 37128 \\
2011 Jan 30 & 2011.0793 & AC-AW-E6 & 17 & 53 & 24760 37128 \\
2011 Feb 05 & 2011.0957 & AC-AE AC-E6 & 19 & 68 & 22192 37128 56537 58715 \\
2011 Feb 07 & 2011.1012 & AC-AE AC-E6 AE-W7 E6-W7 & 18 & 76 & 22192 37128 56537 58715 \\
2011 Feb 10 & 2011.1094 & AC-AE-W7 AC-E6-W7 & 19 & 73 & 22192 37128 56537 58715 \\
2011 Feb 11 & 2011.1121 & AC-AE AC-E6 AE-W7 & 18 & 66 & 22192 37128 56537 58715 \\
2011 Feb 12 & 2011.1148 & AC-AE-W7 AC-E6-W7 & 18 & 79 & 22192 37128 56537 58715 \\
2011 Feb 13 & 2011.1175 & AC-E6 E6-W7 & 19 & 79 & 22192 37128 56537 58715 \\
2011 Mar 13 & 2011.1941 & E6-AC & 17 & 47 & 37128 56537 58715 \\
2011 Mar 16 & 2011.2022 & AW-AC E6-AC & 18 & 49 & 37128 56537 58715 \\
2011 Dec 10 & 2011.9393 & AE-AW-E6 W7-AW & 16 & 74 & 23408 23862 37128 56537 \\
2012 Jan 04 & 2012.0075 & AE-AW E6-AW W7-AW & 15 & 70 & 23408 23862 37128 56537 \\
2013 Jan 17 & 2013.0451 & AC-AW-E6 & 17 & 53 & 58142 24760 37128 50019 \\
2013 Feb 01 & 2013.0861 & AC-AW-E6 & 17 & 53 & 22192 24760 37128 \\
\enddata
\tablecomments{The second column is median Julian Year at the time of the observations.}
\end{deluxetable}

\clearpage

\begin{deluxetable}{llcrrccc}
\tabletypesize{\scriptsize}
\tablecaption{NPOI Calibrators. \label{NPOI_table_cal}}
\tablehead{
\colhead{HD}&
\colhead{Spectral}&
\colhead{$V$}&
\colhead{$V-K$}&
\colhead{$E(B-V)$}&
\colhead{$\theta_{V-K}$}&
\colhead{$V^2_{\rm min}$}&
\colhead{Nights} \\
\colhead{}&
\colhead{Classification}&
\colhead{(mag)}&
\colhead{(mag)}&
\colhead{(mag)}&
\colhead{(mas)}&
\colhead{}&
\colhead{}
}
\startdata
HD    886&B2IV& 2.83&-0.94& 0.01&0.50&0.97& 2\\
HD   5448&A5V& 3.87& 0.23&0&0.70&0.82& 1\\
HD  11171&F3III& 4.65& 0.78&0&0.65&0.89& 1\\
HD  17573&B8Vn& 3.63&-0.23& 0.01&0.55&0.96& 3\\
HD  19994&F8V& 5.06& 1.31& 0.05&0.75&0.84& 1\\
HD  21790&B9Vs& 4.74&-0.15&0&0.34&0.96& 1\\
HD  22192&B5Ve& 4.23& 0.12& 0.11&0.51&0.90& 5\\
HD  23408&B8III& 3.87&-0.12& 0.04&0.50&0.92& 2\\
HD  23862&B7p& 5.09& 0.15& 0.05&0.35&0.96& 2\\
HD  24760&B0.5V& 2.88&-0.83& 0.11&0.52&0.90& 5\\
HD  25940&B3Ve& 4.04& 0.24& 0.17&0.61&0.86& 1\\
HD  37128&B0Ia& 1.70&-0.57& 0.04&1.01&0.45&16\\
HD  50019&A3III& 3.60& 0.44& 0.02&0.80&0.96& 1\\
HD  56537&A3V& 3.57& 0.03& 0.02&0.67&0.72& 5\\
HD  58142&A1V& 4.64& 0.07&0&0.41&0.99& 1\\
HD  58715&B8Vvar& 2.90&-0.20& 0.02&0.78&0.63& 3\\
HD  67006&A2V& 4.84& 0.18& 0&0.41&0.95& 1\\
HD  87504&B9III-IV& 4.60&-0.13&0&0.37&0.99& 1\\
HD  87737&A0Ib& 3.49& 0.19&0&0.75&0.76& 1\\
HD  87887&A0III& 4.49& 0.04&0&0.44&0.99& 1\\
HD  91316&B1Ib& 3.85&-0.43& 0.05&0.42&0.92& 1\\
HD  97633&A2V& 3.34& 0.26&0&0.86&0.78& 4\\
HD  98058&A7IVn& 4.50& 0.37&0&0.54&0.98& 1\\
HD  98664&B9.5Vs& 4.06&-0.08&0&0.49&0.99& 1\\
HD 112413&A0pSiEuHg& 2.90&-0.24&0&0.76&0.92& 4\\
HD 126129&A0V& 5.12& 0.05& 0&0.33&0.99& 1\\
HD 129174&B9p& 4.91&-0.14& 0&0.32&0.99& 1\\
HD 130109&A0V& 3.72& 0.07& 0.01&0.64&0.74& 3\\
HD 193432&B9.5V& 4.76&-0.05&0&0.37&0.98& 1\\
HD 214923&B8.5V& 3.40&-0.17& 0&0.63&0.95& 2\\
HD 219688&B5Vn& 4.40&-0.36& 0.01&0.35&0.95& 1\\
HD 222173&B8V& 4.30&-0.15& 0&0.42&0.95& 1\\
HD 222603&A7V& 4.50& 0.44& 0&0.56&0.99& 1\\
\enddata
\tablecomments{The angular diameter $\theta_{V-K}$ was corrected for extinction. 
$V^2_{\rm min}$ is the minimum estimated calibrator visibility based on $\theta_{V-K}$.}
\end{deluxetable}

\clearpage

\begin{deluxetable}{lcccrrrrrrr}
\tabletypesize{\scriptsize}
\tablecaption{NPOI Positions for $\sigma$ Orionis A,B.\label{NPOI_table_abc}}
\tablehead{
\colhead{}&
\colhead{}&
\colhead{Number of} &
\colhead{$\rho$}&
\colhead{$\theta$}&
\colhead{$\sigma_{\rm maj}$}&
\colhead{$\sigma_{\rm min}$}&
\colhead{$\phi$}&
\colhead{O--C$_\rho$}&
\colhead{O--C$_\theta$}
\\
\colhead{UT Date}&
\colhead{Julian Year}&
\colhead{visibilities}&
\colhead{(mas)}&
\colhead{(deg)}&
\colhead{(mas)}&
\colhead{(mas)}&
\colhead{(deg)}&
\colhead{(mas)}&
\colhead{(deg)}&
\\
\colhead{(1)}&
\colhead{(2)}&
\colhead{(3)}&
\colhead{(4)}&
\colhead{(5)}&
\colhead{(6)}&
\colhead{(7)}&
\colhead{(8)}&
\colhead{(9)}&
\colhead{(10)}
}
\startdata
2001 Feb 21\dotfill & 2001.1412 &  276 & 255.61 & 109.42 &  3.10 &  0.97 & 165.4 & -0.74 &  -0.5 \\
2001 Feb 22\dotfill & 2001.1440 &  184 & 256.43 & 109.80 &  2.02 &  0.63 & 164.1 &  0.08 &  -0.1 \\
2002 Feb 15\dotfill & 2002.1241 &   45 & 256.36 & 107.01 &  6.72 &  1.24 & 168.4 & -0.18 &  -0.5 \\
2006 Nov 08\dotfill & 2006.8524 &  121 & 256.55 &  96.40 &  0.86 &  0.65 & 112.4 &  0.17 &  -0.1 \\
2006 Nov 18\dotfill & 2006.8798 &  234 & 256.13 &  96.13 &  2.06 &  0.72 & 166.7 & -0.20 &  -0.3 \\
2007 Feb 18\dotfill & 2007.1316 &   94 & 255.20 &  95.52 &  0.59 &  0.42 & 165.1 & -0.95 &  -0.2 \\
2007 Oct 19\dotfill & 2007.7969 &   75 & 256.73 &  94.08 &  1.47 &  0.66 &  69.2 &  0.61 &  -0.2 \\
2007 Oct 20\dotfill & 2007.7997 &   45 & 255.83 &  94.09 &  1.31 &  0.66 &  64.6 & -0.29 &  -0.2 \\
2007 Nov 01\dotfill & 2007.8325 &   92 & 255.76 &  94.12 &  2.08 &  0.73 &  11.1 & -0.34 &  -0.0 \\
2007 Nov 03\dotfill & 2007.8380 &  136 & 255.85 &  94.06 &  1.79 &  0.65 &   7.4 & -0.25 &  -0.1 \\
2008 Nov 17\dotfill & 2008.8784 &  373 & 257.09 &  91.15 &  4.96 &  2.25 &  18.9 &  0.87 &  -0.6 \\
2010 Mar 18\dotfill & 2010.2090 & 1320 & 254.39 &  88.44 &  4.26 &  0.85 &  96.8 & -1.69 &  -0.1 \\
2010 Mar 21\dotfill & 2010.2172 & 1143 & 257.09 &  88.48 &  2.52 &  0.48 &  92.8 &  1.01 &  -0.1 \\
2010 Mar 25\dotfill & 2010.2282 & 1455 & 255.08 &  88.37 &  2.11 &  0.40 & 117.6 & -1.01 &  -0.1 \\
2010 Mar 29\dotfill & 2010.2391 &  495 & 256.70 &  88.36 &  2.50 &  0.40 &  90.0 &  0.59 &  -0.1 \\
2011 Jan 29\dotfill & 2011.0769 &  668 & 256.90 &  86.52 &  2.67 &  0.83 &  11.1 &  0.59 &   0.1 \\
2011 Jan 30\dotfill & 2011.0796 &  633 & 256.98 &  86.58 &  2.48 &  0.81 &   9.2 &  0.64 &   0.2 \\
2011 Feb 05\dotfill & 2011.0961 &   58 & 259.63 &  85.26 &  6.66 &  0.90 &  16.7 &  3.16 &  -1.1 \\
2011 Feb 10\dotfill & 2011.1097 &  391 & 256.57 &  87.06 &  3.88 &  1.25 &  11.5 &  0.10 &   0.7 \\
2011 Feb 11\dotfill & 2011.1125 &  207 & 256.18 &  87.43 &  3.38 &  0.76 &  15.0 & -0.28 &   1.0 \\
2011 Feb 12\dotfill & 2011.1152 & 1479 & 257.02 &  86.81 &  7.59 &  1.28 &  17.7 &  0.56 &   0.4 \\
2011 Feb 13\dotfill & 2011.1180 &  348 & 256.79 &  86.91 &  2.00 &  0.58 & 170.3 &  0.34 &   0.5 \\
2011 Mar 13\dotfill & 2011.1946 &   30 & 260.14 &  84.73 &  5.44 &  0.89 &  16.0 &  3.87 &  -1.6 \\
2011 Mar 16\dotfill & 2011.2028 &   58 & 257.54 &  86.38 &  2.65 &  0.64 &  14.1 &  1.29 &   0.1 \\
2011 Dec 10\dotfill & 2011.9393 &  772 & 256.83 &  84.17 &  7.29 &  1.31 & 175.3 &  0.47 &  -0.4 \\
\enddata
\tablecomments{The second column is Julian Year at 7 UT on the date of observation.}
\end{deluxetable}

\clearpage

\begin{deluxetable}{lcccrrrrrrr}
\tabletypesize{\scriptsize}
\tablecaption{NPOI Positions for $\sigma$ Orionis Aa,Ab\label{NPOI_table_ab}}
\tablehead{
\colhead{}&
\colhead{HJD $-$}&
\colhead{Number of} &
\colhead{$\rho$}&
\colhead{$\theta$}&
\colhead{$\sigma_{\rm maj}$}&
\colhead{$\sigma_{\rm min}$}&
\colhead{$\phi$}&
\colhead{O--C$_\rho$}&
\colhead{O--C$_\theta$}
\\
\colhead{UT Date}&
\colhead{$2,400,000$}&
\colhead{visibilities}&
\colhead{(mas)}&
\colhead{(deg)}&
\colhead{(mas)}&
\colhead{(mas)}&
\colhead{(deg)}&
\colhead{(mas)}&
\colhead{(deg)}&
\\
\colhead{(1)}&
\colhead{(2)}&
\colhead{(3)}&
\colhead{(4)}&
\colhead{(5)}&
\colhead{(6)}&
\colhead{(7)}&
\colhead{(8)}&
\colhead{(9)}&
\colhead{(10)}
}
\startdata
2006 Nov 18 & 54057.79 & 234&  6.72& 186.56&  0.38&  0.12&152.6&  0.33&  -0.7\\
2007 Feb 18 & 54149.79 &  94&  2.12& 229.95&  0.42&  0.13&127.7& -0.16&  -9.8\\
2007 Oct 19 & 54392.79 &  75&  6.93& 203.80&  0.90&  0.18& 49.4&  0.10&   1.1\\
2007 Nov 01 & 54405.79 &  92&  5.91& 208.04&  0.53&  0.22&179.4& -0.08&   0.3\\
2007 Nov 03 & 54407.79 & 136&  5.42& 209.85&  0.69&  0.26&  9.5& -0.41&   1.2\\
2008 Nov 16 & 54786.79 & 901&  7.38& 191.22&  0.31&  0.13&153.2&  0.28&  -0.6\\
2010 Mar 18 & 55273.79 &1320&  5.18& 212.36&  0.08&  0.06&135.0& -0.01&   0.1\\
2010 Mar 25 & 55280.79 &1455&  4.35& 217.89&  0.09&  0.08&135.0& -0.05&   0.7\\
2011 Jan 29 & 55590.79 & 668&  1.01& 309.77&  0.30&  0.08&  0.1& -0.05&   2.8\\
2011 Jan 30 & 55591.79 & 633&  1.03& 326.23&  0.47&  0.09&  6.3& -0.00&   4.5\\
2011 Feb 07 & 55599.79 & 270&  1.10& 123.74&  0.53&  0.16&  2.4&  0.03&  -7.7\\
2011 Feb 11 & 55603.79 & 207&  3.05& 169.08&  0.76&  0.18& 11.7&  0.92&   9.8\\
2011 Feb 12 & 55604.79 &1479&  2.58& 163.92&  0.20&  0.09&  2.1&  0.20&   1.5\\
2011 Dec 10 & 55905.79 & 772&  4.73& 181.47&  0.34&  0.10&167.7& -0.28&   0.9\\
2012 Jan 04 & 55930.79 &  86&  7.15& 190.50&  0.45&  0.14&170.5&  0.11&  -0.8\\
2013 Jan 17 & 56309.79 & 450&  0.49& 318.20&  0.72&  0.13&  3.7& -0.53& -33.9\\
2013 Feb 01 & 56324.79 & 555&  3.44& 170.90&  0.58&  0.10& 12.1&  0.17&   0.3\\
\enddata
\tablecomments{HJD is computed at local midnight (7 UT) on the date of observation.}
\end{deluxetable}

\begin{deluxetable}{lrrrrrr} 
\tablewidth{0pt}
\tablecaption{CTIO radial velocities measured for $\sigma$ Orionis Aa,Ab\label{tab.rv}} 
\tablehead{ 
\colhead{}        & \colhead{HJD $-$}   & \colhead{}      & \colhead{$V_r$ (Aa)} & \colhead{$V_r$ (Ab)} & \colhead{$O-C_{\rm Aa}$} & \colhead{$O-C_{\rm Ab}$} \\
\colhead{UT Date} & \colhead{2,400,000} & \colhead{Phase} & \colhead{(km\,s$^{-1}$)} & \colhead{(km\,s$^{-1}$)} & \colhead{(km\,s$^{-1}$)} & \colhead{(km\,s$^{-1}$)}  }
\startdata 
2008 Sep 23 & 54732.804 &  0.977 &  -46.11 $\pm$  15.9 &   163.1 $\pm$   8.1 &    8.3 &    3.6 \\
2008 Sep 23 & 54732.886 &  0.978 &  -54.28 $\pm$  17.2 &   166.6 $\pm$   9.4 &    1.3 &    5.5 \\
2008 Sep 24 & 54733.796 &  0.984 &  -65.71 $\pm$  16.3 &   187.5 $\pm$   9.1 &    4.1 &    7.6 \\
2008 Sep 24 & 54733.878 &  0.985 &  -71.51 $\pm$  15.8 &   188.2 $\pm$   8.4 &   -0.4 &    6.6 \\
2008 Sep 25 & 54734.801 &  0.991 &  -87.81 $\pm$  15.6 &   203.4 $\pm$   8.5 &   -4.1 &    5.0 \\
2008 Sep 25 & 54734.889 &  0.992 &  -84.40 $\pm$  16.7 &   203.8 $\pm$   8.7 &    0.1 &    4.3 \\
2008 Sep 26 & 54735.857 &  0.999 &  -80.21 $\pm$  16.9 &   204.0 $\pm$   9.0 &    4.7 &    4.0 \\
2008 Sep 26 & 54735.896 &  0.999 &  -84.20 $\pm$  16.4 &   204.0 $\pm$   9.5 &    0.3 &    4.5 \\
2008 Sep 27 & 54736.794 &  0.005 &  -64.47 $\pm$  16.2 &   180.6 $\pm$   9.0 &    3.0 &    3.8 \\
2008 Sep 27 & 54736.877 &  0.006 &  -61.04 $\pm$  17.1 &   177.2 $\pm$   9.4 &    4.2 &    3.2 \\
2008 Sep 28 & 54737.789 &  0.012 &  -33.41 $\pm$  19.5 &   145.1 $\pm$   9.5 &    6.7 &    4.6 \\
2008 Sep 28 & 54737.872 &  0.013 &  -31.65 $\pm$  21.1 &   141.2 $\pm$  10.5 &    6.2 &    3.6 \\
2008 Sep 29 & 54738.795 &  0.019 &    6.17 $\pm$  27.2 &   108.9 $\pm$  11.4 &   22.1 &    0.4 \\
2008 Sep 29 & 54738.879 &  0.020 &    2.51 $\pm$  26.9 &   105.9 $\pm$  10.2 &   16.7 &   -0.3 \\
2008 Sep 30 & 54739.779 &  0.026 &   13.86 $\pm$  25.1 &    85.5 $\pm$  10.8 &   12.4 &    0.0 \\
2008 Sep 30 & 54739.882 &  0.027 &   15.48 $\pm$  23.5 &    82.6 $\pm$  10.5 &   12.5 &   -1.0 \\
2008 Oct 02 & 54741.853 &  0.041 &   27.53 $\pm$  18.1 &    56.8 $\pm$   9.9 &    4.0 &    0.5 \\
2008 Oct 02 & 54741.874 &  0.041 &   24.18 $\pm$  23.1 &    58.8 $\pm$   9.7 &    0.5 &    2.7 \\
2008 Oct 16 & 54755.837 &  0.138 &   51.46 $\pm$  22.6 &    14.5 $\pm$  11.5 &   -2.6 &   -1.3 \\
2008 Nov 14 & 54784.806 &  0.340 &   59.26 $\pm$  24.4 &    14.0 $\pm$  10.7 &    3.0 &    1.2 \\
2008 Dec 16 & 54816.817 &  0.564 &   51.84 $\pm$  25.3 &    18.7 $\pm$  11.2 &    1.4 &   -2.0 \\
2008 Dec 18 & 54818.848 &  0.578 &   53.32 $\pm$  24.1 &    21.8 $\pm$  10.0 &    3.5 &    0.4 \\
2008 Dec 23 & 54823.841 &  0.613 &   50.61 $\pm$  21.1 &    22.6 $\pm$  11.3 &    2.2 &   -0.8 \\
2008 Dec 24 & 54824.756 &  0.619 &   54.19 $\pm$  29.1 &    24.7 $\pm$  10.4 &    6.1 &    0.9 \\
2009 Jan 10 & 54841.757 &  0.738 &   40.18 $\pm$  17.0 &    33.2 $\pm$  10.0 &   -0.6 &   -0.2 \\
2009 Jan 16 & 54847.760 &  0.780 &   36.93 $\pm$  15.0 &    38.4 $\pm$  11.4 &    0.0 &   -0.1 \\
2009 Jan 21 & 54852.701 &  0.815 &   41.04 $\pm$  27.8 &    37.9 $\pm$  10.4 &    8.2 &   -6.0 \\
2009 Jan 21 & 54852.729 &  0.815 &   54.09 $\pm$  50.6 &    37.3 $\pm$   9.0 &   21.3 &   -6.7 \\
2009 Jan 30 & 54861.712 &  0.878 &   26.73 $\pm$  21.8 &    58.7 $\pm$  10.8 &    5.6 &   -0.7 \\
2009 Feb 02 & 54864.585 &  0.898 &   24.12 $\pm$  21.0 &    68.3 $\pm$  11.0 &    8.8 &    1.1 \\
2009 Feb 03 & 54865.629 &  0.905 &   18.61 $\pm$  22.2 &    69.8 $\pm$  10.0 &    5.9 &   -0.8 \\
2009 Feb 04 & 54866.592 &  0.912 &   17.39 $\pm$  22.0 &    73.1 $\pm$   9.9 &    7.3 &   -1.0 \\
2009 Feb 10 & 54872.520 &  0.953 &    3.01 $\pm$  12.2 &   110.4 $\pm$   5.7 &   20.3 &    0.1 \\
2009 Feb 13 & 54875.537 &  0.974 &  -40.70 $\pm$  17.8 &   157.5 $\pm$   8.8 &    7.0 &    6.8 \\
2009 Feb 13 & 54875.661 &  0.975 &  -42.83 $\pm$  18.3 &   158.6 $\pm$   9.2 &    6.6 &    5.6 \\
2009 Feb 17 & 54879.643 &  0.003 &  -72.54 $\pm$  16.4 &   191.0 $\pm$   9.0 &    3.2 &    3.2 \\
2009 Feb 18 & 54880.666 &  0.010 &  -43.96 $\pm$  15.1 &   160.7 $\pm$   7.7 &    5.0 &    8.4 \\
2009 Feb 19 & 54881.664 &  0.017 &  -24.64 $\pm$  25.4 &   121.0 $\pm$  17.0 &   -1.5 &    2.9 \\
2009 Feb 20 & 54882.626 &  0.024 &   18.37 $\pm$  22.5 &    91.6 $\pm$  11.6 &   22.5 &   -1.3 \\
2009 Feb 21 & 54883.634 &  0.031 &   20.08 $\pm$  21.6 &    71.2 $\pm$  10.1 &    9.9 &   -2.8 \\
2012 Nov 04 & 56235.782 &  0.473 &   47.41 $\pm$   3.9 &    13.3 $\pm$   1.9 &   -6.0 &   -3.3 \\
2013 Jan 07 & 56299.742 &  0.920 &   17.72 $\pm$   7.9 &    79.2 $\pm$   2.7 &   11.3 &    0.3 \\
2013 Jan 13 & 56305.696 &  0.961 &   -2.82 $\pm$   4.3 &   125.9 $\pm$   2.6 &   24.1 &    2.7 \\
2013 Jan 15 & 56307.693 &  0.975 &  -48.15 $\pm$   9.1 &   156.2 $\pm$   3.6 &    1.8 &    2.5 \\
2013 Jan 19 & 56311.594 &  0.002 &  -75.28 $\pm$   5.9 &   189.3 $\pm$   3.1 &    1.4 &    0.3 \\
2013 Jan 22 & 56314.664 &  0.024 &   -3.32 $\pm$   6.2 &    91.3 $\pm$   2.5 &    0.1 &   -0.6 \\
2013 Jan 25 & 56317.652 &  0.045 &   27.46 $\pm$   2.6 &    51.6 $\pm$   2.5 &   -0.2 &    0.9 \\
2013 Jan 28 & 56320.674 &  0.066 &   38.59 $\pm$   4.7 &    33.0 $\pm$   2.3 &   -2.1 &   -0.5 \\
2013 Jan 31 & 56323.667 &  0.087 &   46.47 $\pm$   6.1 &    21.8 $\pm$   2.2 &   -0.8 &   -3.0 \\
2013 Feb 02 & 56325.607 &  0.100 &   47.96 $\pm$   4.4 &    19.0 $\pm$   2.0 &   -1.9 &   -2.3 \\
2016 Jan 21 & 57408.586 &  0.663 &   46.30 $\pm$   6.2 &    23.2 $\pm$   1.8 &    0.5 &   -3.5 \\
2016 Feb 14 & 57432.602 &  0.831 &   32.63 $\pm$   3.7 &    47.5 $\pm$   2.9 &    2.2 &    0.4 \\
2016 Mar 02 & 57449.593 &  0.949 &   -9.07 $\pm$   9.7 &   105.3 $\pm$   2.7 &    4.6 &   -0.2 \\
2016 Mar 03 & 57450.630 &  0.957 &  -25.75 $\pm$   6.0 &   116.7 $\pm$   2.2 &   -4.5 &    1.1 \\
2016 Mar 11 & 57458.571 &  0.012 &  -36.12 $\pm$  10.1 &   143.7 $\pm$   2.9 &    4.3 &    2.7 \\
2016 Mar 12 & 57459.546 &  0.019 &  -19.76 $\pm$   7.3 &   110.3 $\pm$   3.0 &   -3.0 &    0.7 \\
2016 Mar 13 & 57460.536 &  0.026 &   18.06 $\pm$   6.8 &    84.6 $\pm$   2.4 &   17.2 &   -1.7 \\
2016 Mar 14 & 57461.553 &  0.033 &   21.06 $\pm$   6.4 &    70.3 $\pm$   3.1 &    7.3 &    1.1 \\
2016 Mar 15 & 57462.559 &  0.040 &   23.98 $\pm$   6.1 &    59.3 $\pm$   3.4 &    1.1 &    2.1 \\
2016 Mar 25 & 57472.526 &  0.110 &   49.28 $\pm$   3.0 &    17.0 $\pm$   2.0 &   -2.0 &   -2.5 \\
2016 Mar 27 & 57475.480 &  0.130 &   51.03 $\pm$   5.8 &    14.8 $\pm$   2.7 &   -2.4 &   -1.8 \\
\enddata
\end{deluxetable}

\clearpage

\begin{deluxetable}{lcccc}    
\tablewidth{0pt}
\tablecaption{Orbital parameters for the close pair $\sigma$ Orionis Aa,Ab} 
\tablehead{
\colhead{}               & \colhead{}                     & \colhead{}         & \colhead{}           & \colhead{Simultaneous Visual and} \\
\colhead{Parameter}      & \colhead{\citet{simondiaz15}}  & \colhead{CTIO RV}  & \colhead{MIRC Only}  & \colhead{Spectroscopic Fit\tablenotemark{a}}}
\startdata 
$P$ (d)                   & 143.198 $\pm$ 0.005    & 143.1995  $\pm$ 0.0031 & 143.224    $\pm$ 0.020   &  143.2002  $\pm$ 0.0024 \\
$T$ (HJD$-$2,400,000)     & 56,597.623 $\pm$ 0.024 & 56,597.605 $\pm$ 0.045 & 56,597.684 $\pm$ 0.061   &  56,597.638 $\pm$ 0.014 \\
$e$                       & 0.7782 $\pm$ 0.0011    & 0.7804    $\pm$ 0.0022 & 0.77845    $\pm$ 0.00083 &  0.77896   $\pm$ 0.00043 \\
$\omega_{\rm Aa}$ ($^\circ$) & 199.98 $\pm$ 0.24      & 199.56    $\pm$ 0.47   & 199.61     $\pm$ 0.46    &  199.83    $\pm$ 0.12  \\
$a$ (mas)                 & \nodata                & \nodata                & 4.2861     $\pm$ 0.0069  &  4.2860    $\pm$ 0.0031 $\pm$ 0.0107 \\
$i$ ($^\circ$)             & \nodata                & \nodata                & 56.48      $\pm$ 0.14    &  56.378    $\pm$ 0.085   \\
$\Omega$ ($^\circ$)        & \nodata                & \nodata                & 7.02       $\pm$ 0.23    &  6.878     $\pm$ 0.079   \\
$K_{\rm Aa}$ (km\,s$^{-1}$) & 71.9 $\pm$ 0.3         & 65.18     $\pm$ 1.26   & \nodata                  &  72.03     $\pm$ 0.25  \\
$K_{\rm Ab}$ (km\,s$^{-1}$) & 95.2 $\pm$ 0.3         & 99.39     $\pm$ 0.89   & \nodata                  &  95.53     $\pm$ 0.22  \\
$\gamma$ (km\,s$^{-1}$)    & 31.10 $\pm$ 0.16       & \nodata                & \nodata                  &  31.18     $\pm$ 0.21  \\
$\gamma_{\rm CTIO}$ (km\,s$^{-1}$)  &  \nodata        & 36.37     $\pm$ 0.38   & \nodata                  &  37.63     $\pm$ 0.35 
\label{tab.orb}
\enddata
\tablenotetext{a}{Based on a simultaneous visual and spectroscopic orbit fit to the positions measured with MIRC at the CHARA Array, radial velocities published by \citet{simondiaz15}, and radial velocities measured at CTIO.  The angle between the ascending node and periastron, as referenced to $\sigma$ Ori Ab, is given by $\omega_{\rm Ab}$ = $\omega_{Aa}$ + 180$^\circ$ = 19\fdg83. The final reduced $\chi^2_\nu$ for the final simultaneous orbit fit is 1.25, with a break down of $\chi^2_{\nu,{\rm SimonDiaz}} = 1.10$, $\chi^2_{\nu,{\rm CTIO}} = 1.54$, and $\chi^2_{\nu,{\rm MIRC}} = 1.60$.} 
\end{deluxetable}

\begin{deluxetable}{lcc}
\tablecaption{Orbital elements of $\sigma$ Orionis A,B\label{NPOI_table_orbit}}
\tablehead{
\colhead{Parameter}&
\colhead{\citet{turner08}} &
\colhead{This work}
}
\startdata
$P$ (days)                  & $57235 \pm 1096$       & $58402 \pm 2$      \\
$T$ (JD)                    & $2451362 \pm 3726$     & $2451255 \pm 39$   \\
$e$                         & $0.0515 \pm 0.0080$    & $0.024 \pm 0.005$ \\
$\omega_{\rm A} (^\circ)$      & $8.7 \pm 16.9$         & $7.4 \pm 9.9$      \\
$a$ (mas)                   & $266.2 \pm 2.1$        & $262.9 \pm 2.2$    \\
$i (^\circ)$                 & $159.7 \pm 3.7$        & $172.1 \pm 4.6$    \\
$\Omega (^\circ)$ (J2000.0)  & $301.7 \pm 9.6$        & $301.6 \pm 10.4$   \\
\enddata
\tablecomments{The angle between the ascending node and periastron, as referenced to $\sigma$ Ori B, is given by $\omega_{\rm B}$ = $\omega_{\rm A}$ + 180$^\circ$ = 187\fdg4.}
\end{deluxetable}

\begin{deluxetable}{lc} 
\tablewidth{0pt}
\tablecaption{Derived Properties for $\sigma$ Orionis Aa, Ab, and B} 
\tablehead{
\colhead{Parameter} & \colhead{Value} }
\startdata 
$M_{Aa}$ ($M_\odot$) & 16.99  $\pm$ 0.20  \\  
$M_{Ab}$ ($M_\odot$) & 12.81  $\pm$ 0.18  \\  
$M_{B}$ ($M_\odot$)  & 11.54  $\pm$ 1.15  \\
$\pi$ (mas)        & 2.5806 $\pm$ 0.0088 \\
$d$ (pc)           & 387.51 $\pm$ 1.32 
\label{tab.properties}
\enddata 
\end{deluxetable}

\end{document}